# Introduction

# A CRYPTOGRAPHIC STUDY OF SOME DIGITAL SIGNATURE SCHEMES

A thesis submitted for the partial fulfillment of the degree of

**Doctor of Philosophy**
in
MATHEMATICS

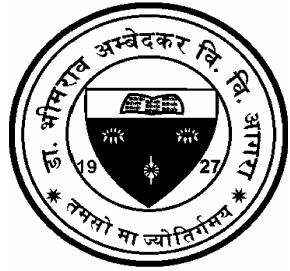

by

**Manoj Kumar**

Under the supervision of

**Dr. Sunder Lal**
Professor & Head,
Department of Mathematics
Institute of Basic Science

Dr. B. R. Ambedker University, Agra

(Formerly Agra University)

2003

*Dedicated*

*To*

*my wife, the most compassionate,*

*kindest person in the universe*

*Chhaya*

*and*

*To my sweet daughter, a medium*

*of worship, Melody of my life*

*Aayushi Raghuvanshi*

# \*\* Declaration\*\*

I do hereby declare the present research work has been carried out by me under the supervision of Prof. Sunder Lal and this work has not been submitted elsewhere for any other degree, diploma, fellowship or any other similar title.

Date: 27 Nov 2003                                              (Manoj Kumar)

                                                                       Lecturer,

                                      Department of Applied Science & Humanities,

                                        Hindustan College of Science & Technology,

                                                              Farah-Mathura-281122.

                                              E Mail: Balyanyamu@rediffmail.com

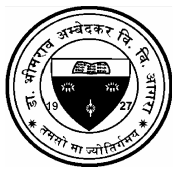

DEPARTMENT OF MATHEMATICS

Institute of Basic Science

Dr. B.R. Ambedkar University

Khandari, Agra-282002

**Sunder Lal**
Prof. & Head

Phone: (R) 2155023
(O) 2152129
email: sunder_lal2@rediffmail.com

# **Certificate**

This is to certify that the thesis entitled **"A CRYPTOGRAPHIC STUDY OF SOME DIGITAL SIGNATURE SCHEMES"** submitted to Dr. B. R. Ambedker University, Agra for the degree of Doctor of Philosophy by **Sri Manoj Kumar,** is a bonafide record of research work done by him under my supervision**.** To the best of my knowledge, this thesis has not previously formed the basis for the award to the candidate of any degree, diploma, fellowship or any other similar title and the work has not been submitted to any university or institution, for the award of any other degree. The thesis presents independent original contribution of the candidate.

**Date: 27 Nov 2003**  (Sunder Lal)

(Supervisor)

# ✷✷Acknowledgment✷✷

I am grateful to my Supervisor Dr. Sunder Lal, Prof. & Head, Department of Mathematics, Institute of Basic Science, Dr. B.R. Ambedkar University, Agra, who spares his valuable time in guiding me for my research work. He encourages me always. I am short in word to express his contribution to this thesis through criticism, suggestions and discussions. My sincere thanks are to Dr. Sanjay Choudhary and Dr. Sanjeev Sharma, both senior lectures in the Department of Mathematics, Institute of Basic Science, Dr. B.R. Ambedkar University, Agra, for their kind suggestions.

I am deeply indebted to Dr. J.P. Arya, HOD, Department of Mathematics, D.A.V. College Muzaffarnagar, who had laid the foundation of my M.Phil. degree and then encouraged me for Ph.D. degree.

There is no word to express my feeling for my family members and relatives, especially to my parent for their hidden cooperation and to my wife Chhaya for her enthusiastic inspirations, round the clock cooperation and help me in many ways. Really, it is not possible to express the love and affection to sweet and little daughter Aayushi Raghuvanshi, who is the driver of my success, to make the path for my Ph.D. work.

I am thankful to all the faculty members and staff of Hindustan College of Science & Technology, Farah – Mathura, Ravindra Kumar (T&P Officer), Jagadeesh. G. (Lecturer, Computer Science), Sri Gopal Sharma, Sri Sarvesh B. Singh for their kind cooperation in this electronic age, through computer, printer etc. I am also thankful to my research colleagues Ms. Meeta Gurmukh, Mr. Atul Churvedi, Mr. Anil Agarwal and Mr. Amit K Awasthi for their result oriented discussion.

At last but not the least, my sincere thanks to the writers of the books and the research papers, which, I have consulted during the course of my research work.

Date: 27 Nov 2003                                                                              (Manoj Kumar

# Chapter 0

# Introduction

# Introduction

*Confidential Communication* is one of the necessities of the social life. In this context, some questions that need attention are:

- *How can one transmit the message secretly, so that no unauthorized person gets knowledge of the message ?*
- *How can the sender ensure himself that the message arrives in the right hands exactly as it was transmitted ?*
- *How can the receiver ensure himself that the message is coming from the right person exactly as it was transmitted?*

Traditionally, there are two ways to solve such problems. One can disguise the very existence of the message, perhaps by writings with invisible ink; or try to transmit the message via a trustworthy person. A different scientific approach to solve these problems is **cryptography**.

```
Cryptography is the art or science of keeping
secrets secret. Cryptography is about secure
communication through insecure channels.
```

The concept of securing messages through cryptography has a long history of at least 4000 years. Ancient Egyptians enciphered some of their hieroglyphic writings on monuments. The first cryptosystem, named as *"SCYTALE"* was established in Greece in the 5th Century B.C. Julius Caesar is credited with using one of the earliest cryptographic systems to send military messages to his Generals. He replaced every A by D, every B by E, and so on through the alphabet. The Greek writer Polybius devised a cryptosystem, in which he replaced letters to numbers. Arabs first wrote proper Cryptology as early as 900 AD and used it extensively in 1412. In 1460, Leon Alberti devised a cipher wheel. In 1585, Blaise De Vigenere described the polyalphabetic substitution cipher. In 1790, Jefferson, developed a cylinder comprised 26 disks, each with a random alphabets. The key was order of disks.

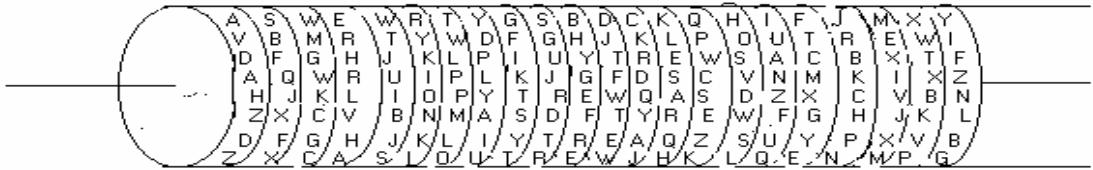

Fig-1: Jefferson Cylinder

In 1817, Wadsworth invented a disc, which was later developed by Wheatstone in 1860 and now known as the *Wheatstone disc*. It generates a polyalphabetic cipher using two concentric wheels. Enigma Rotor machine, one of important class of machines, heavily used during World War II, comprised a series of rotor wheels with internal cross- connections, providing a substitution using a continuously alphabet.

After the First World War, things begin to change. U.S. Army and Navy organizations, working secretly and confidentially, began to make fundamental advances in cryptography. During the 30s and 40s, a few basic papers did appear in the public literature and several treatises on the subject were published, notable among these are the basic papers by Hill [46] on matrix-based cryptosystems. Claude Shannon published his two fundamental papers about communication theory. In 1948, **A Mathematical Theory of Communication** [99] and in 1949, **Communication Theory of Secret Systems** [100], appeared in the Bell System Technical Journal.

From 1949 to 1967, the cryptographic literature was not in abundance. In that year a different sort of contribution appeared: David Kahn's history, **The Codebreakers.** It did not contain any new technical ideas but it contains a remarkably complete history of what had gone before, including mention of something that the U.S. government still consider secret. Codebreaker enjoyed a good sale and made tens of thousands of people, who had never given the matter a moment's thought, aware of cryptography. A trickle of new cryptographic papers began to be written.

Horst Feistel, who had earlier worked on identification friend or foe devices for the Air Force, took his life long passion for cryptography to IBM Watson Laboratory in Yorktown Height, New York. There he began the development of what was to become the U.S. *Data Encryption Standard (DES)*. By the early 1970s, several technical reports on this subject by Feistel and his colleagues had been made public by IBM.

In 1976, *Whitfield Diffie and Martin Hellman* presented their famous paper, **New Direction in Cryptography** [24]. This paper altered the face of cryptography. The result has been a spectacular increase in the number of people working in cryptography, the number of meetings held, and the number of books and papers published.

## 0.1. Classical Cryptosystems

The classical goal of cryptography is privacy: two party (sender S and receiver R) wish to communicate privately, so that an adversary knows nothing about what was communicated. A standard cryptographic solution to the privacy problem is *classical cryptosystems (Symmetric cryptosystems)*. Classical cryptosystems provide secure communication for a pair of user. In classical key cryptosystem, the encryption key can be calculated from the decryption key and vice versa. In most cryptosystems, the encryption key and decryption key are the same. These cryptosystems, also called **secret key cryptosystems, single key cryptosystems,** or **one key cryptosystems,** require that the sender and the receiver agree on a key before they can communicate securely. The security of the classical key cryptosystem rests in the key; divulging the key means, that anyone can encrypt and decrypt the messages. As long as the key remains secret, the communication remains secret.

Classical Cryptosystem consists of the following:

- A plaintext space $\mathcal{P}$ of all possible plaintexts P.
- A ciphertext space $\mathcal{C}$ of all possible cryptotexts C.
- A key space $\mathcal{K}$ in which each K determines an encryption algorithm $E_K$ and a decryption algorithm $D_K$ such that $D_K(E_K(P)) = P; P \in \mathcal{P}$.

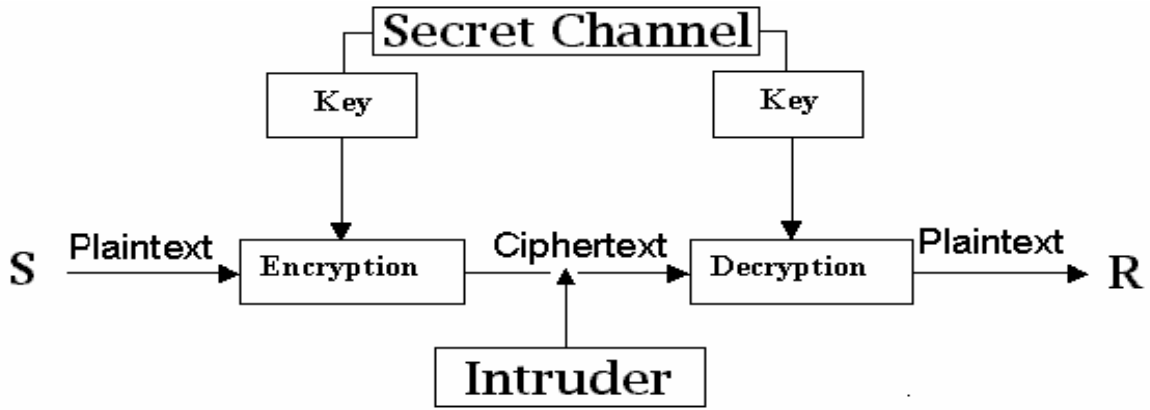

Fig - 2 : Classical Cryptosystem

Using symmetric cryptosystem, it is safe to send encrypted message without fear of interception because an interceptor is unlikely to be able to decipher the message; however, with this advantage, symmetric cryptosystems also have the following problems:

- *Sender and receiver share a common secret key, which must be transmitted before an encryption procedure can start.*
  - *There is always a problem of how to transfer the common secret key. How do you send the secret key from the sender to the recipient securely ? If you could send the secret key securely, then, in theory, you would not need the symmetric cryptosystem for your communication ?*
  - *Classical cryptosystems are unable to deal the dispute between the sender and receiver, as to what message, if any, was sent. In classical cryptosystems, third party cannot verify who actually send the message. Because the secret key is common, however the receiver has the ability to produce any ciphertext that could have been produced by the sender.*
  - *Key management is the major problem in classical cryptosystem. In classical cryptosystems, every pair of users*

*has to have an exclusive secret key; n users need n. (n-1)/2 different keys. The following fig shows the key management of the six users A, B, C, D, E and F in classical cryptosystem.*

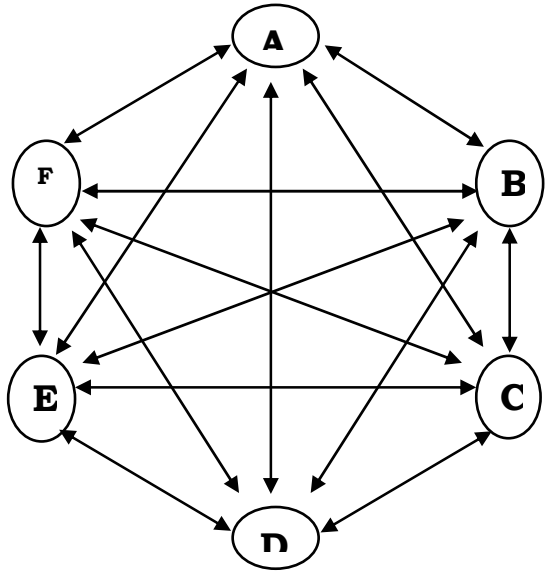

**Fig-3: Key Management in Classical Cryptosystem**

- *We can achieve authentication and message integrity by using classical cryptosystems, but No one other than the receiver can check the message integrity and authenticity of the message.*

0.2. Public Key Cryptosystems

Keeping in view the above drawbacks of Classical Cryptosystems and to answer all the question, **Whitefield Diffie** and **Martin Hellman** proposed in 1976, a new type of cryptosystem which is called now, as ***public key cryptosystem* [PKC]**. The invention of public key cryptography is the most important event in the field of cryptography and provided answers to all the above problems of key managements and digital signatures.

A public key cryptosystem is a pair of families $\{E_K\}$ and $\{D_K\}$, $K \in$ key space $K$, of algorithms representing invertible transformations

$$E_K: P \rightarrow C \text{ and } D_K: C \rightarrow P$$

- for every $K$, $E_K$ is the inverse of $D_K$.

- for every $K$, it is easy to compute $E_K$ and $D_K$

- for almost every $K$, each easily computed algorithm equivalent to $D_K$ is computationally infeasible to drive from $E_K$.

- for every $K$, it is feasible to compute inverse pairs $E_K$ and $D_K$ from $K$.

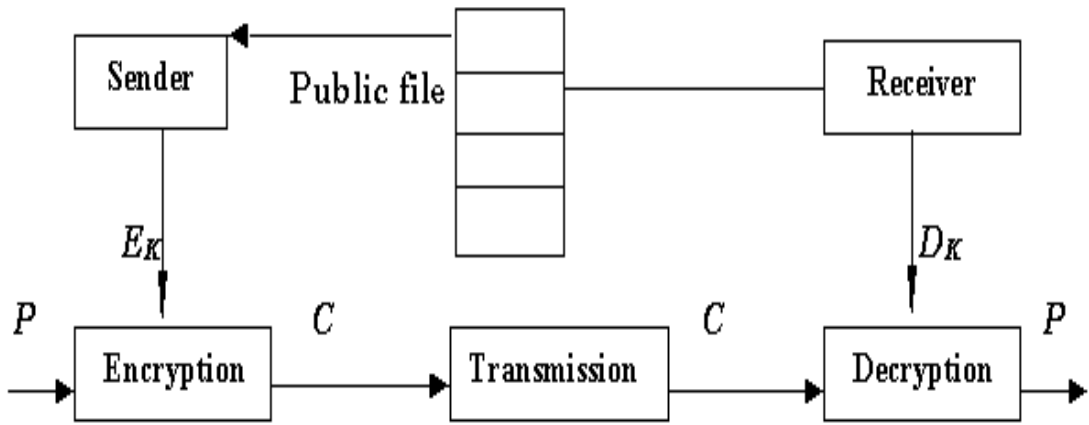

Fig-4: Public Key Cryptosystem.

Because of the third property, the encryption key $E_K$ of each user can be made public without compromising the security of his private decryption key $D_K$. The public key cryptographic system is therefore split into two parts, a family of encryption transformations and a family of decryption transformations in such a way that, given a member of one family, it is infeasible to find the corresponding member of the other.

The forth property guarantees that there is a feasible way of computing corresponding pairs of inverse transformations when no constraint is placed on what either the encryption or decryption transformation is to be.

In this cryptosystem, there is no need to send the secret key via secure channel. Every user A in the system has one pair of keys, a **public key** $E_{KA}$ for encryption and a **private key** $D_{KA}$ for decryption. With the help of private key, the user calculates his public key. It is computationally infeasible to recover the private key by using his public key. At first glance, it seems incredible that such cryptosystems could actually exit. Later we shall see that there are indeed successful realizations. The **RSA** cryptosystem and the **ElGamal** cryptosystems are such realizations. All public keys are publicly available; they might be stored in public file, as in a phone book. On the other hands, the private keys are kept secret; only the owner knows them. Let us now suppose that we have a public key cryptosystem. Consider the user A wants to send the message *m* to B. The process works as follows.

- A looks up the public key $E_{KB}$ of B, encrypts the message *m* as $E_{KB}(m) = C$ and send the cryptotext C to B.

- B is able to decrypt the cryptotext C, since he exclusively knows the key $D_{KB}$. He gets $m = D_{KB}(E_{KB}(m))$.

- No other user can decipher $E_{KB}(m)$, since no one can recover $D_{KB}$ from $E_{KB}$ and $E_{KB}(m)$.

Public Key Cryptosystem has the following noteworthy advantages.

- *No secret key exchange among the users.*
- *Every user needs relatively few keys.*
- *New user can join the system without disturbing the old users.*
- *Encryption and decryption.*
- *Communication in a secure way over the open channels for peoples who have never met.*
- *Digital signature schemes.*
- *Nonrepudiation.*

Some well-known public key cryptosystems are described below:

## RSA Cryptosystem

In 1977, three people who were to make the single most spectacular contribution to the public key cryptography: Ronald Rivest, Adi Shamir and Leonard Adleman…took up the challenge of producing a full-fledged public key cryptosystem, which is called as RSA cryptosystem. The process lasted several months during which Rivest proposed approaches, Adleman attacked them and Shamir recalls doing some of each. In may 1977 they were rewarded with success…they have discovered how a simple piece of classical number theory could be made to solve the problems [25] ?

RSA cryptosystem is based upon the difficulty of factorization of two large primes. In this cryptosystem, we assume there is a key center. The key center

- Generates two large distinct prime $p$ and $q$, each roughly the same size.
- Calculates $n = p.q$ and $\phi(n) = (p-1).(q-1)$.
- For every user, selects randomly integer e; $1 \leq e \leq \phi(n)$ *i.e.* $(e, \phi(n)) = 1$.
- Computes $d$, the multiplicative inverse of e.mod $\phi(n)$, by Euclidean algorithm.

The pair $(e, n)$ is the public key or the encryption key and the pair $(d, n)$ is the secret key or the decryption key of the user A. The encryption transformation is a mapping $f$ from $Z/Z_n$ to itself given by $C = f(P) = P^e \mod n$. The decryption transformation is given by $P = f^{-1}(C) = C^d \mod n$.

If a user B encrypts a message $m$ for the user A, then B should do the following:

- Represents the message as an integer in the interval $[0, n-1]$
- Computes $C = m^e \mod n$.
- Sends the ciphertext C to the user A.
- A recovers the message $m$ by computing $C^d \mod n$, using his secret key $d$.

## ElGamal Cryptosystem

Taher ElGamal [28] proposed a cryptosystem based on the difficulty of finding **discrete logarithm**. This difficulty of discrete logarithm is applicable not only to the key management, but also to the design of encryption and signature algorithm. To go further, it is necessary to understand the concept of discrete logarithm. Suppose we have a finite field $Z_q$ with group operation of multiplication. With the help of repeated square method anyone can compute $b^x$ for large x such as $y = b^x$, y and b $\in Z_q$. Now it is easy to compute $y = b^x$ for given x and b but for a given b and y, it becomes infeasible to compute x. Here the value x is said to be the discrete logarithm of y to the base b. For example, let us take $Z_{19}$ and $b = 2$, then we can compute $2^6 = 7$, which is equivalent to saying that the **discrete logarithm of 7 to the base 2 is equal to** 6. But it is much more difficult to compute the number 6 by using 2 and 7. So for a large x, it is computationally infeasible to recover this x with the known b and y.

In the ElGamal scheme there is a general agreement upon a prime $p$ and an integer g. Every user A select randomly an integer $\alpha_A$; $0 < \alpha_A < p - 1$ and compute a public value $\beta = g^{\alpha_A} \mod p$. If we want to send a message $m$ to the user A then we choose random integer $k$ and send to A the pair ($g^k$, $m\beta^k$). Now since the user A knows $\alpha_A$ so he can recover the message $m$ from this pair. The user A raises $\alpha_A^{th}$ power to the first element of the pair and then divide the second element of the pair to get the plaintext.

## Messey - Omura Cryptosystem

This cryptosystem is also based on the difficulty of finding discrete logarithms. All the users have agreed upon a public prime $p$. Now each user A chooses two positive integers $e_A$ and $d_A$ such that $e_A d_A \mod p = 1$

*In contrast with RSA cryptosystem, in this system the users keep both the numbers secret, publishing neither of them.* Now consider the situation in which the user A sends the message $m$ to the user B. We assume that a number represents the message, which is less than $p$.

The algorithms works as follows:

- The user A computes $m^{e_A} \mod p$ and sends to the user B.

- The user B computes the $e_B^{th}$ power of the number he has received and return the result $m^{e_A e_B} \mod p$ to the user A.

- Now the user A applies his number $d_A$ to what he received and gets $m^{e_A e_B d_A} \mod p$. This number turns out to be $m^{e_B} \mod p$. The user A sends this result to the user B.

- The user B applies $d_B$ to the received number and obtains the message $m$.

## Knapsack Cryptosystem

In 1978, Merkle and Hellman [71] proposed a public key cryptosystem, which is based upon the so-called Knapsack problem. Given a set of $\{v_i\}$ of $k$ positive integers and an integer $V$, find a $k$-bit integer $n = (\varepsilon_{k-1} \varepsilon_{k-2}............... \varepsilon_1 \varepsilon_0)_2$, where $\varepsilon_i \in \{0,1\}$ are the binary digit of $n$ such that

$$\sum_{i=0}^{k-1} \varepsilon_i v_i = V,\text{ if such an } n \text{ exists.}$$

A special case of the Knapsack problem is the superincreasing Knapsack problem. *In this sequence the terms arranged in increasing order, have the additional property that each term is greater than the sum of all the earlier one*. To solve the superincreasing Knapsack problem is a very easy task. We observe the given $v_i$, starting with the largest, until we get to the first one that is $\leq V$. The following algorithm solves the Knapsack problem for a given superincreasing sequence of $k$-tuple and integer $V$:

- Set $W = V$ and $j = k$.

- Starting with $\varepsilon_{j-1}$ and decreasing the index of $\varepsilon$, choose all the $\varepsilon_i$ equal to zero until we get the first $i$, call it $i_0$, such that $v_{i_0} \leq W$. Set $\varepsilon_{i_0} = 1$.

- Replace $W$ by $W - v_{i_0}$, set $j = i_0$, and $W > 0$, go back to step 2.

- If $W = 0$, you are done. If $W > 0$ and all the remaining $v_i > W$, then there is no solution to the given problem.

## Construction of Knapsack Cryptosystem

Each user chooses at random a super increasing $k$-tuple $\{v_0, v_1, v_2... v_{k-1}\}$, an integer $m$ which is greater than $\sum_{i=0}^{k-1} v_i$, and an integer $a$ prime to $m$, $0 < a < m$. For example, we could choose an arbitrary sequence of $k+1$ positive integer $z_i$, $i = 0,1,2,3...,k$, less than some convenient bound; set $v_0 = z_0$, $v_i = z_i + v_{i-1} + v_{i-2} +....+ v_0$ for $i = 0,1,2,3....,k-1$; and set $m$ equal to $z_k + \sum_{i=0}^{k-1} v_i$. Then one can chooses a random positive integer $a_0 < m$ and take $a$ to be the first integer $\geq a_0$ that is prime to $m$. After that, one computes the multiplicative inverse of $a$ mod $m$ = $b$ and also compute the $k$-tuple $\{w_i\}$ defined by $w_i = a.v_i \mod m$. The user keeps the numbers $v_i, m, a$ and $b$ but publishes the $k$-tuple of $w_i$. The encryption key is $K_E = \{w_0, w_1 ..... w_{k-1}\}$. The decryption key is $K_D = \{b, m\}$.

Someone who wants to send a plaintext $k$- bits message $P = (\varepsilon_{k-1} \varepsilon_{k-2} ..... \varepsilon_1 \varepsilon_0)_2$ to a user with enciphering key $\{w_i\}$, he computes $C = f(P) = \sum_{i=0}^{k-1} \varepsilon_i w_i$, and transmits the integer. To read the message, the receiver first the least positive residue $V$ of $bC$ mod $m$. Since $bC \equiv \sum_{i=0}^{k-1} b\varepsilon_i w_i \equiv \sum_{i=0}^{k-1} \varepsilon_i v_i$, it follows that $V = \sum_{i=0}^{k-1} \varepsilon_i v_i$. Here we are using the fact that both $V < m$ and $\sum_{i=0}^{k-1} \varepsilon_i v_i \leq \sum_{i=0}^{k-1} v_i < m$ to convert the congruence modulo $m$ to equality.

## Quadratic Residue Cryptosystem

This Cryptosystem was proposed by Blum Blum and Shub [6] in 1986. The Quadratic Residue of an integer b mod N is simply the residue b$^2$ mod N; if 'a' is a quadratic residue modulo N, it means that there exists an integer which gives 'a' as remainder when the square of the integer is calculated under modulo N.

To explain the working of QRC, consider two users R and S. R is the recipient and S is the sender of the message. In QRC, each user 'R' chooses two large prime $p$ and $q$, subject to them both being congruent to 3 mod 4. These two primes become the secret key and $n = p.q$ the public key of the user R. The public key is made available and anyone wishing to send R a message proceeds as follows.

The sender S chooses any quadric residue in the $Z_n^*$. This residue is becomes the seed $x_0$ for the key generator, which will be used to encrypt the message. The sender S adopts some prearranged scheme to assign binary values to the residues, such as writing 1, when the parity is even, and 0, when the parity is odd. To proceed, S writes down the appropriate key bit according to his first residue $x_0$, and then squares this residue. S then writes down the next corresponding bit, and square the resulting residue again, and so on. In this manner, pseudorandom binary bit pattern is produced which can be used as a one time pad to encrypt the message, providing of course, that the seed is not reused with the modulus. The significance of the one time pad is that such a cipher is the only one providing complete theoretical security.

The legitimate recipient R can easily recover the plaintext, because knowing the two factors of $n = pq$, he can compute two integer a and b such that $ap + bq = 1$. He can compute the residue $x_{i-1}$ by taking the square root of the residue $x_i$, then the square root of $x_i$ and so on right back to the residue $x_0$. On receipt of the message, knowing its length $N + 1$ and the final residue $x_N$, R computes the following values to recover the seed $x_0$ and then the plaintext.

$$\mu = [(p + 1)/4]^N \mod p–1,$$
$$\theta = [(q + 1)/4]^N \mod q–1,$$
$$u = (x_N \mod p)^\mu \mod p,$$
$$v = (x_N \mod q)^\theta \mod q,$$
$$x_0 = (bqu + apv) \mod n.$$

## Hybrid Cryptosystem: Diffie Hellman's Key Exchange System

The implementation of public key cryptosystems is slow. While, there are some good and very fast classical cryptosystems available. It makes sense to try combination of the two types of cryptosystems, exploiting each system advantages. This is the fundamental idea of a hybrid cryptosystem given by Diffie and Hellman [24].

Essentially, one uses a symmetric cryptosystem for communication. A public key cryptosystem is used only for the exchange the secret of keys, necessary for the symmetric cryptosystem.

In order to send the secret key $k$ to the user B, one must encrypt $k$ using B's public key. Then the user B decrypts the received value with his private key and obtains $k$. This procedure works perfectly; it has only one disadvantage, if the user A sends the secret key to the user B, in order that they can communicate. From an abstracts point view, both users should agree upon a key, rather than one of them choosing a key and sending it to the other. Using such a key generation both users would play exactly the same role.

Consider two users A and B who wish to agree upon a secret key and who have already agreed upon two things, a prime $p$ and a positive integer $g < p$, both of which may be public. So for instance, all the users could use the same $p$ and g. The system works as follows:

- The user A chooses a secret integer a $< p$ and compute $\alpha = g^a \mod p$. He sends $\alpha$ to the user B.

- The user B chooses a secret integer b $< p$ and compute $\beta = g^b \mod p$. He sends $\beta$ to the user A.

- The user A computes $k = \beta^a \mod p$ and B computes $k = \alpha^b \mod p$.

In this way, both the user have agreed upon a common secret key $k$. This system is secure against an attack of any interceptor. The interceptor knows $\alpha$ and $\beta$. If he deduces 'a' from $\alpha$ or 'b' from $\beta$ then he could compute $k$ just as the user A and B did. However, it is infeasible to deduce these values. This is one of the basic properties of discrete logarithm.

Already we have listed the noteworthy advantages of public key cryptography over classical cryptography. Among these advantages, digital signature is an important cryptographic tool, which provides secrecy and authenticity for the digital communication.

## 0.3. Digital Signatures

We use Physical signature in common practice for authentication. As we move to the world where our decisions and agreements are communicated electronically, we need to replicate these procedures. Public key cryptography (***PKC***) provides mechanisms for such procedures through digital signature schemes. In order to establish the authenticity of the electronic message, before a message is sent out, the sender of the message would sign it using a digital signature scheme (DSS) and then encrypt the message, using encryption algorithm. Thus, a digital signature scheme proves the authenticity of the message as well as the authenticity of the sender.

A *digital signature scheme* is one of the most important cryptographic tools in ***PKC***, essential in various security services. Digital signatures have the same role for the digital message that the handwritten signatures have for documents on paper. A digital signature scheme consists of the followings:

- A signature generation algorithm, which is a method for producing a digital signature.

- A signature verification algorithm, which is a method for verifying a digital signature.

The crucial difference from a hand written signature is that the *digital signature is a data string which intimately connected with the message with some originating entity, whereas the handwritten signature adjoined to the message and always looks the same. Consequently, no one can alter the digital signature without everyone being able to see that the document has been changed. Thus, the digital signature ensures the integrity of the signed document.* For, the digital signature is altered at all, then the application of the public key to the altered signed message yields a plaintext, which will appear very random. *This property of the digital signature implies that the signature is not reusable and unforgeable.* Nobody, except the signer can sign a document. Hence, the recipient is convinced that the signer deliberately signed the document. The digital signature ensures the *authenticity* of the signer. Finally, these two properties *(unforgeablity and authenticity)* ensure the *non- repudiation* of the signature. The signer cannot deny later that he did not sign the message.

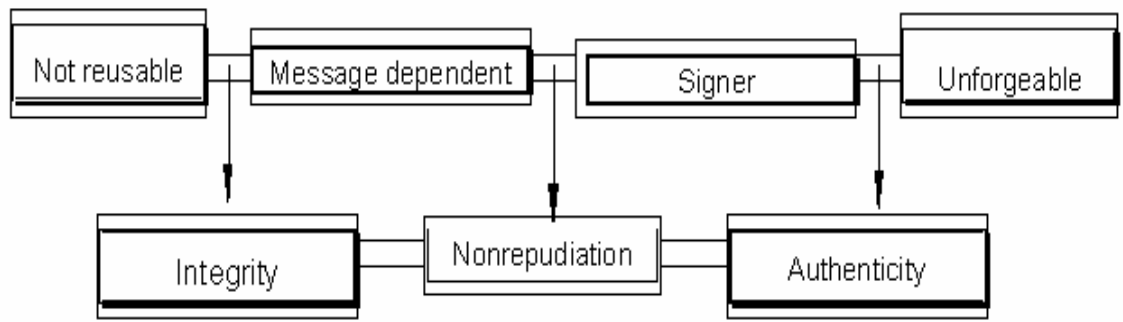

**Fig-5: Properties of Digital Signature**

## A Classification of Digital Signature Schemes

There are two classes of digital signature schemes, *digital signature schemes with appendix* require the original message as input to the verification algorithm and digital signature schemes with message recovery do not require the original message as input to the verification algorithm.

## Digital Signature Schemes with Appendix

Digital signature schemes with appendix rely on cryptographic hash function $h$. DSA, ElGamal and Schnorr signature scheme are such digital signature schemes. A priori knowledge of the message is required for the verification algorithm. In these signature schemes, a user A can produces a signature $s^* \in S$ (a set of elements called signature space) for a message $m \in M$ (a set of elements called message space), which can later be verified by any user B.

Each user A selects a secret key $S_A$ from the set $\{S_{A,k} : k \in R\}$. Here $S_{A,k}$ is a one – one mapping from $M_h$ to $S$ and called a signing transformation and $R$ is a set of elements called the indexing set of signing. $M_h$ is the image of $h$ i.e. $h: M \rightarrow M_h$ ; $M_h \subseteq M_S$ ( a set of elements called signing space) is called the hash value space. $S_A$ defines a corresponding mapping $V_A$ from $M_h \times S$ to {true, false} such that

$$V_A(m^*, s^*) = \begin{cases} true, if\ S_{A,k}(m^*) = s^*, \\ flase, otherwise \end{cases}$$

for all $m^* \in M_h$, $s^* \in S$; here, $m^* = h(m)$ for $m \in M$. $V_A$, the public key of the signer is also called a verification transformation and is constructed such that it may be computed without knowledge of the secret key $S_A$ of the signer.

To generate the signature,

- A selects his secret key $S_A$ and an element $k \in R$ and computes $m^* = h(m)$ and $s^* = S_{A,k}(m^*)$. Here $h$ is a one-way hash function.

- The pair $(s^*, m)$ is the signature on the message $m$, send to user B.

To verify the signature,

- The user B obtains the public key of the user A and computes $m^* = h(m)$ and $u = V_A(m^*, s^*)$.

- The user B accepts the signature if and only if $u = true$.

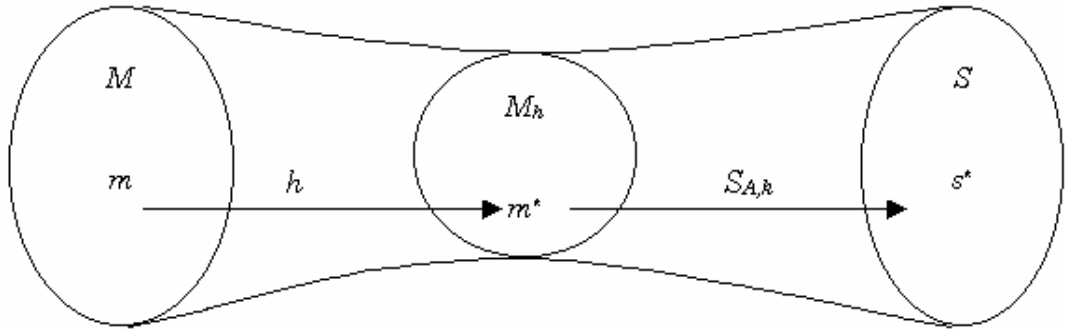

**Fig- 6: The Signing Process**

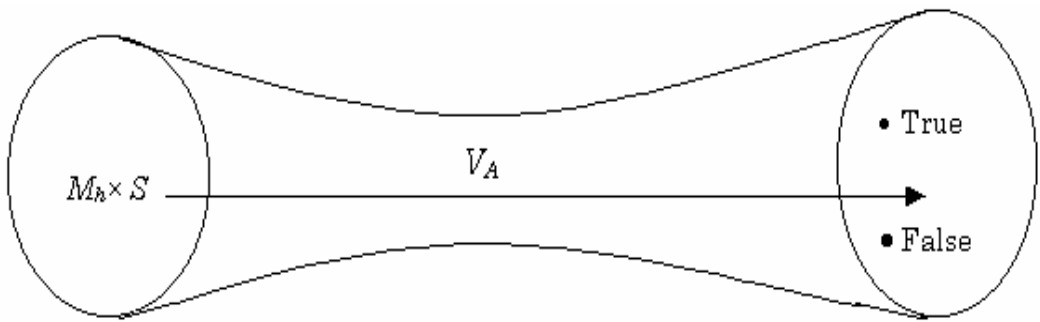

**Fig- 7: The Signature Verification Process**

### Remarks:

- For the signing process $k \in R$ and $S_{A,k}$ should be efficient to compute.

- For the verification process, $V_A$ should be efficient to compute.

- It should be computationally infeasible for any user other A to find a message $m \in M$ and $s^* \in S$ such that

    $V_A (m^*, s^*)$ = true, where $m^* = h(m)$.

- The one – way hash function $h$ is typically selected to be a collision – free hash function.

## Digital Signature Schemes with Message Recovery

Digital signatures with message recovery have the feature that the message can be recovered from the signature itself. These schemes are useful for short messages. A priori knowledge of the message is not required for the verification algorithm. The RSA, Rabin and Nyberg – Rueppel are such digital signature schemes.

To generate the signature,

- A selects his secret key $S_A$ and an element $k \in R$ and computes $m^* = \xi(m)$ and $s^* = S_{A,k}(m^*)$. Here $\xi$ is a one-one mapping from $M$ to $M_S$.

- The number $s^*$ is the signature on the message $m$, which is made available to the user B.

To verify the signature,

- The user B obtains the public key of the user A and computes $m^* = V_A(s^*)$ and verifies $m^* \in M_\xi$ (the image of $\xi$). If $m^* \notin M_\xi$, then reject the signature.

- The user B recovers the message $m$ from $m^*$ by computing $\xi^{-1}(m^*)$.

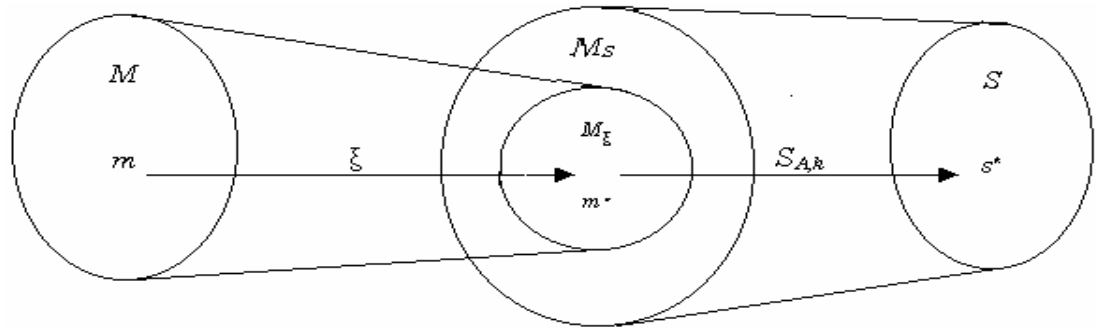

**Fig- 8: Digital Signature with Message Recovery.**

## Remarks:

- Most digital signature schemes with message recovery are applied to messages of a fixed length, while digital signature schemes with appendix are applied to messages of arbitrary length.

- The function $\xi$ is publicly known and $\xi^{-1}$ is easy to compute.

- Any digital signature scheme with message recovery can be converted into a digital signature scheme with appendix by simply hashing the message and than signing the hash value

## The RSA Signature Scheme

The RSA signature scheme was the first digital signature scheme with message recovery. RSA digital signature scheme is one of the most practical and versatile techniques available. The message space, signing space, signature space and ciphertext space for this scheme belong to $Z_n = \{0,1,2...n-1\}$ where $n = p.q$ is the product of two randomly chosen large prime numbers.

To signs a message $m \in M$,

- Any user A computes $m^* = \xi(m)$.

- Applying his secret key $d$, he computes $s^* = (m^*)^d \mod n$.

- The number $s^*$ is the signature of the user A for the message $m$.

To verify the signature $s^*$ and to recover the message $m$,

- Any user B applying the public key $e$ of the user A computes

$$m^* = (s^*)^e \mod n.$$

- Verify that $m^* \in M_\xi$; if not reject the signature.

- B recovers the message $m = \xi^{-1}(m^*)$.

## Feige-Fiat- Shamir Signature Scheme

Feige- Fiat- Shamir signature scheme is digital signature scheme with appendix. This scheme required a one – way hash function $h: \{0, 1\}^* \to \{0, 1\}^k$ for some fixed positive integer $k$. Here $\{0, 1\}^k$ denotes the set of bit- strings of bit-length $k$ and $\{0, 1\}^*$ denotes the set of all bit-strings of any bit-lengths. In this signature scheme,

- Each user A selects a positive integer $k$ and distinct random integers $s_j \in Z_n^*$, where $j = 1,2,...k$ and $n = p.q$.
- A computes $v_j = s_j^{-2} \mod n$.
- The $k$-tuple $(v_j)$ is the public key and the $k$-tuple $(s_j)$ is the secret key of the user A.

To generate the signature for a message $m$,

- A selects a random number $r \in Z_n^*$ and computes $u = r^2 \bmod n$.
- A computes $e = (e_j) = h(m \| u)$; each $e_j \in \{0, 1\}$.
- A computes $s = r \cdot \prod_{j=1}^{k} s_j^{e_j} \bmod n$.
- A's signature for the message $m$ is $(e, s, m)$.

To verify the signature $(e, s, m)$, Any user B

- computes $w = s^2 \cdot \prod_{j=1}^{k} v_j^{e_j} \bmod n$.
- computes $\check{e} = h(m \| u)$.
- accepts the signature if and only if $e = \check{e}$

Unlike the RSA signature scheme, all users may use the same modulus $n = p.q$. In this scenario, a key distribution center would need to generate the primes $p$ and $q$ and the secret and public keys for each user.

## ElGamal Digital Signature Scheme

In August 1991, the U.S. National Institute of Standard and Technology (NIST) proposed a digital signature algorithm (DSA). The DSA has becomes a U.S. Federal Information Processing Standard (FIPS 186) called the Digital Signature Standard (DSS). DSS is the first digital signature scheme recognized by any government. DSA is a variant of ElGamal signature scheme. To go further, first we shall discuss the *ElGamal signature scheme* and then **DSA** and *Schnorr signature scheme* as its variants.

The ElGamal signature scheme is a signature scheme with appendix. It requires a hash function $h: \{0, 1\}^* \to Z_p$, where $p$ is large prime. In this signature scheme, each user A selects,

- a large prime modulous $p$,
- a number $g$ is the generator of $Z_p^*$).
- a collision free one-way hash function $h$;

- a random integer $x_A$, $1 \leq x_A \leq p-2$ and computes $y_A = g^{x_A} \mod p$.

Here $x_A$ is the private key and $y_A$ is the public key of the user A.

To signs a binary message $m$ of arbitrary length, the user A,

- selects a random secret integer $k$, $1 \leq k \leq p-2$, with $(k, p-1) = 1$.

- computes $r = g^k \mod p$ and $k^{-1} \mod p-1$.

- computes $s = k^{-1}\{h(m) - x_A \cdot r\} \mod p-1$.

- A's signature for the message $m$ is $(r, s, m)$.

To verify the signature $(r, s, m)$, Any user B,

- verifies that $1 \leq r \leq p-1$; if not, then rejects the signature.

- computes $v_1 = y_A^r \cdot r^s \mod p$.

- computes $h(m)$ and $v_2 = g^{h(m)} \mod p$.

- accepts the signature if and only if $v_1 = v_2$.

## The Digital Signature Algorithm (DSA)

For this algorithm, we need the following parameters.

1. $p$ = a prime modulus, L bits longs where $512 \leq L \leq 1024$ and L is multiple of 64.

2. $q$ = a 160 bits prime divisor of $p-1$.

3. $g = a^{(p-1)/q} \mod p$, where $a \in Z_p^*$ and g in element of order $q$ in $Z_p^*$.

4. $x_A$ an integer with $0 < x_A < q$.

5. $y_A = g^{x_A} \mod p$.

6. $h$, a one way hash function.

The system parameters, *p, q, h* and *g* are public. The user's public key and the private key are $y_A$ and $x_A$ respectively. To creates a signature for a binary message of arbitrary length, the user A,

- selects a random *k* and computes $k^{-1}$ mod *q*.

- computes $r = (g^k \bmod p) \bmod q$.

- computes $s = k^{-1}\{h(m) + x_A.r\} \bmod q$.

The pair (*r, s*) is the signature of the user A for the message *m*. A new value of *k* must be generated for each new signature.

To verify the signature, the recipient B

- verifies that $1 \leq r \leq p-1$; and $1 \leq s \leq p-1$; if not, then rejects the signature.

- computes $w = s^{-1} \bmod p$ and $h(m)$.

- computes $u_1 = w.h(m) \bmod q$ and $u_2 = rw \bmod q$.

- computes $v = g^{u_1} y_A^{u_2} \bmod q$.

- accepts the signature if and only if $v = r$.

- The security of the DSA relies on two distinct but related discrete

- ogarithm problems. One is the logarithm problem in $Z_p^*$; the other is the logarithm problem in the cyclic subgroup of order *q*.

## The Schnorr Signature Scheme

The Schnorr signature scheme is a well-known variant of the ElGamal signature scheme. This signature also requires a hash function *h*: {0, 1} → $Z_q$ .In this scheme, to signs a binary message *m* of any length, the user A,

- chooses at random secret integer $k_A \in Z_q$ .

- computes $r_A = h(g^{k_A} \bmod p, m)$ and $S_A = k_A - x_A.r_A \bmod p$.

The pair $(r_A, S_A)$ is the signature of the user A for the message *m*. To verify the signature, the recipient B checks the equality

$$r_A = h(g^{S_A} \ y^{r_A} \bmod p, m).$$

## The ElGamal Scheme with Message Recovery

The ElGamal signature scheme and its variants are digital signature scheme with appendix. In contrast, ***Nyberg- Rueppel signature*** is a variant of ElGamal signature scheme, which has the feature that the message can be recovered from the signature itself.

## Nyberg- Rueppel Digital Signature Scheme

The key generation for the Nyberg- Rueppel signature is the same as DSA key generation and the signing space is $M_S = Z_p^*$, $p$ is prime and the signature space is $S = Z_p \times Z_q$, $q$ is a prime and $q$ divides $(p-1)$.

To signs a message $m$, any user A

- computes $m^* = \xi(m)$.

- selects a random secret integer $k \in Z_q^*$ and computes $r = g^{-k} \bmod p$.

- computes $e = m^* r \bmod p$ and $s = e \cdot x_A + k \bmod q$.

The pair $(e, s)$ is the signature of the user A for the message.

To verify the signature, any user B

- verify that $e \in Z_p^*$ and $s \in Z_q^*$; if not, reject the signature.

- computes $v = g^s \ y_A^{-e} \bmod p$ and $m^* = v \cdot e \bmod p$.

- verify that $m^* \in M_\xi$; If $m^* \notin M_\xi$, then reject the signature.

- recovers the message $m$ from $m^*$ by computing $\xi^{-1}(m^*)$.

## Digital Signature with Additional Functionality

There are many alternative digital signature schemes, which provide functionality of a situation beyond authentication and other properties. In these digital signature schemes, a basic digital signature scheme (e.g., The RSA, The ElGamal) is combined with a specific protocol to achieve the additional features which the basic method does not provide

According to the additional features, particular types of the realization of the digital signature works. Some well-known digital signature schemes with additional functionality are described below:

## Multi-Signature Scheme

In many commercial applications, the signature of more than one person is required on a document. When more than one key is required for signature, we call this signature as multi-signature [40,41,43,44,48,80,81]. These signatures are useful when a company issued cheques, which need to be authorized by more than one person. These signatures are also useful in case of contracts, which are to be signed by their business partners.

## Group Signature Scheme

Consider a group of people, in which every member of the group is authorized to sign the documents on behalf of the group. This type of signature is called as group signature [15,18]. Group signature allows its members to sign message in such a way that

- Only members of the group can sign the messages.
- The receiver of the signature can verify that it is a valid group signature.
- The receiver of the signature cannot identify which member of the group is the signer.
- In case of disputes, later on either the group members together or a trusted authority can identify the signer.

## Threshold Signature Scheme

Usually the signer of a digital signature scheme is a single user. Nevertheless, there are situations when a group takes decision and the message is on behalf of an organization. A common example of this policy is a large bank transaction, which requires signatures from more than one person. Such a policy could be implemented by having a separate digital signature for every required signer, but this solution increases the efforts to verify the message linearly with the signer. An alternative method is a threshold signature scheme [20,21,22,23,31,47,50,90], which is based on the **threshold cryptosystems**. In a threshold scheme, a secret is shared among a group of people in such a manner that a certain number of them can work together to recover the secret. This distribution of key provides protection against dishonest group members and accidental disclosure of secret. To go further first we shall discuss a threshold scheme due to Shamir.

A $(t, n)$ threshold scheme is a scheme to distribute a secret key K into $n$ users in such a way that any $t$ users can cooperate to reconstruct K but a collusion of $t-1$ or less users reveal nothing about the secret. Shamir's scheme is based on Lagrange interpolation in a field. To implement it, a polynomial $f$ of degree $t-1$ is randomly chosen in $Zq$ such that $f(0) = K$. Each user $i$ is given a public identity $u_i$ and a secret share $f(u_i)$. Now any $t$ out of $n$ shareholders can reconstruct the secret K = $f(0)$, by pooling their shares and using

$$f(0) = \sum_{i=1}^{t} f(u_i) \prod_{j=1, j\neq i}^{t} \frac{-u_j}{u_i - u_j} \bmod q$$

Here we assume for simplicity that the authorized subset of $t$ users consists of shareholders $i$ for $i=1,2,3...t$.

In $(t, n)$ threshold signature scheme, any subgroup of $t$ or more shareholders of the designated group can generate a valid group signature in such a way that the verifier can check the validity of the signature without identifying the identities of the signers. Unfortunately, threshold schemes suffer from a drawback. In $(t, n)$ threshold schemes, when any $t$ or more shareholders act in collusion, they can impersonate any other set of shareholders to forge the signatures. In this case, the malicious set of signers does not have any responsibility for the signatures and it is impossible to trace the signers.

In general, there are some requirements that (t, n) threshold signatures should satisfy:

1. The group secret key '*K*' can be divided into 'n' different "secret shares", $k_1$, $k_2$, $k_3$... $k_n$, such that

    a. The group signatures can be easily produced with knowledge of any 't' secret shares and $t \leq n$

    b. With knowledge of any $(t - 1)$ or fewer secret shares, it is impossible to generate a group signature.

    c. The group secret or any secret share cannot be derived from the released group signature and all partial signatures.

2. It is better that the size of the group signature is equivalent to the size of an individual signature.

3. Any outsider can verify the group signature and verification process should be as simple as possible.

4. The signing groups hold the responsibility to signed message. That is, each user in this group cannot deny having sign the message.

5. The partial signature and group signature cannot be forged by any subset of malicious shareholders.

## Undeniable Signature Scheme

Digital signatures unlike handwritten signature and banknote printing are easily copied exactly. This property can be advantageous for some uses, such as dissemination of the announcements and public keys, where the more copies distributed better. Nevertheless, it is unsuitable for many other applications. Consider electronic replacements for all the written or oral commitments that are to some extant personally or commercially sensitive. In such cases, the proliferation of certified copies could facilitate improper uses like blackmail or industrial espionage. The recipient of such a commitment should be able to ensure that the issuer cannot later disavow it, but the recipient should also be unable to show that the commitment to anyone else without the issuer's consent.

For the above situation, undeniable signatures [7,13,14,16,42,83] are well suited. An undeniable signature, like a digital signature, is a number issued by a signer that depends on the message issued. Unlike a digital signature, an undeniable signature can only be verified with the help of the signer.

## Blind Signature Scheme

Rather than basic digital signature scheme, blind signature schemes [10,12,27,59,110] are two – party protocols between the sender A and the signer B. The basic idea is that the sender A sends a piece of information (a message $m$) to signer B which B signs and returns to A. For this signature, A can compute B's signature on a priori message $m$ of his choice. At the completion of the protocol, B knows neither the message $m$ nor the signature associated with it. The purpose of blind signature is to prevent the signer B from observing the message it signs and the signature; hence, it is later unable to associate the signed message with the sender A.

Blind signature schemes have applications where the sender A (the customer) does not want the signer B (the bank) to be capable of associating a postiori a message $m$ and a signature $S_B(m)$ to a specific instance of the protocol. This may be important in electronic cash applications where a message $m$ might represent a monetary value that A can spend. When $m$ and $S_B(m)$ are presented to B for payment, B is unable to deduce which party was originally given the signed value. This allows A to remain anonymous so that spending patterns cannot be monitored. A blind signature scheme protocol requires the following components:

- A digital signature scheme for the signer B.
- Functions $f$ and $\zeta$ (known only to sender A) such that

$$\zeta(S_B(f(m))) = S_B(m).$$

Function $f$ is called a blinding function, $\zeta$ an unblinding function and $f(m)$ a blind message.

## Proxy Signature Scheme

Delegation of rights is a common practice in the real world. If a manager of a company goes on holiday, he has to delegate to his deputy the capability to sign on behalf of the company. In the paper based world a corporate seal is used for this purpose. A corporate seal represents the organization, not the person who has authority to use the seal.

In the electronic world to facilitate this requirement, proxy signature scheme [48,63,69,102,105,109,111,112] has been invented to delegate signing capability efficiently and transparently. Digital signature schemes rely on a secret signature key, which only the certified person knows. If this secret key is delegated to another person directly, it can no longer be identified with that person and, hence the assumption of the digital signature is broken. We therefore need to delegate the signing capability without revealing the secret key such that the recipient can verify the signature of the original signer with the help of proxy signer (a person who can create the signature on behalf of original signer).

## Directed Signature Scheme

In many situations, signed message is sensitive to the signature receiver. Signatures on medical records, tax information and most personal/business transactions are such situations. Consider when a user A wants to generate a signature on a message *m,* sensitive for B and the message is also of concern to other users. For this situation, the form of the signature should be such that only B can directly verify the signature and that B can prove its validity to any third party C, whenever necessary. Such signatures are called **directed signatures** [56,57,67]. In directed signature scheme, the signature receiver B has full control over the signature verification process. Nobody can check the validity of signature without his cooperation.

## Abstract of the Thesis

In this thesis, *we propose some directed signature schemes.* In addition, we have discussed their applications in different situations. In this thesis, we would like to discuss the security aspects during the design process of the proposed directed digital signature schemes. The security of the most digital signature schemes widely use in practice is based on the two difficult problems, viz; the problem of factoring integers (The RSA scheme) and the problem of finding discrete logarithms over finite fields (The ElGamal scheme).

The proposed works in this thesis is divided into seven chapters. The next discussion provides the abstract of each chapter.

## Chapter 1

In this chapter, we propose ***a directed signature scheme*** and discuss some of its applications. To generate the signature, the signer uses the public key of the signature receiver in such a way that at the time of signature verification, the secret key of the signature receiver is requires. As a result, only the signature receiver can check the validity of signature. The signature receiver is able to prove the validity of the signature to any third whenever necessary.

Next, an application to threshold verification is discussed and proposed ***a directed signature scheme with threshold verification***. This scheme is applicable when message is transmitted for an organization. A signer can generate a signature on the message $m$ to the organization $R$ in such a way that any subset $H_R$ of $k$ or more users of a designated group $G_R$ of $n$ users can determine the validity of the signature but any $k-1$ or less users cannot verify the signature. In this scheme, there is no share distribution center (**SDC**). The security parameters and secret shares are not fixed and the signer can change the threshold value. The signer is free to decide the signing policy.

Next, one more to threshold cryptosystem is discussed. Threshold cryptosystems are useful when the decision is taken on behalf of an organization. The organization has a single public key. The secret key is distributed among the users of organization in such a way that any $k$ or more users required to reads the message, but when any $k-1$ or less users works together, they cannot receive any information about the message.

## Chapter 2

Consider a situation when the message is sensitive to the signature receiver 'C' and an original signer 'A' wants to delegate his signing power to his proxy signer 'B'. The proxy signer creates the signature on behalf of the original signer. For illustration, consider a situation, where, there is a NGO's hospital, which facilitates AIDS checkup. The hospital is headed by a CMO A, who has many responsibilities to perform. He designates a trusted colleague B to issue certificate on his behalf. Any patient C does not want to disclose the result of his/her check-up and he/she wants a certificate that nobody can read without his help. Unfortunately, if C is HIV positive then there is need of curing. For treatment there is another important hospital headed by a NGO chief Y, Where all types of Medicine, remedies and all resources are available for AIDS patients C. For treatment there, C has to prove the validity of a certificate to a desired person. For this situation, in this chapter, we combine the idea of proxy signature scheme with directed signature scheme and propose a directed delegated signature scheme.

In this signature scheme the signing of message is done in two phases. The first phase is off-line. This phase is governed by proxy signature scheme, and can be performed even before the message to be signed is known. In this phase, the original signer delivers his signing power to his proxy signer. In this scheme, the signer A does not delegate his secret key directly to his proxy signer B. He delegates his signing right (secret key) to his proxy signer through a proxy signature key. Without knowing the secret key of the original signer, the proxy signer can generate the signature on behalf of original signer.

The second phase is on-line. It starts after the message become known and utilizes the precomputation of the first phase. In this phase, the proxy signer generates the directed signature on behalf of original signer.

To verify the signature, the secret key of the signature receiver is required. As a result, no one other than the signature receiver can check the validity of the signature. In case of third party verification, the signature receiver is able to prove the validity of the signature through zero knowledge protocol.

## Chapter 3

Already we have discussed the role of threshold signature scheme. In many situations, signed message is sensitive to the signature receiver and the signer is generally a single person. However, when the message is on behalf of an organization, a valid message may require the approval or consent of several people. For this situation, we propose a ***directed threshold signature scheme.*** In this signature scheme, let $G$ be a group of $n$ designated users, out of which any $t$ or more members can generate the signature on a message *m* for a user B. The user B can verify the signature and that B can prove its validity to any third party C, whenever necessary. Nobody can check the validity of the signature without the help of B.

For our construction, we assume the existence of a trusted share distribution center (SDC) which determines the group secrets parameters and the secret shares $v_i$, $i \in G$. Let *H* be any subset of G, containing *t* members. We also assume the existence of a designated combiner *DC* who collects partial signatures from each user in the subgroup *H*. Every shareholder in the group has equal authority with regard to the group secret. The generation of the directed signature requires *t* out of *n* shareholders and interaction with *DC*. We should note that there is no secret information associated with the *DC*.

## Chapter 4

In most situations, the signer and the verifier is generally a single person. However, when the message is sent by one organization to another organization, a valid message may require the approval or consent of several people. In this case, the signature generation and verification is done by more than one consenting rather than by a single person. This chapter proposes a ***threshold signature scheme with threshold verification.***

Consider the message is transmitted by an organization *S* to another organization *R* and may require the approval of a subset $H_S$ of *t* or more signer from a designated group $G_S$ of *n* users belongs to the organization *S*. On the other hand, the signing group wants to generate the signature on a message *m* in such a way that the signature can be verified by any subset $H_R$ of *k* or more users from a designated group $G_R$ of *l* users belongs to the organization *R*, then threshold verification schemes serve our purpose.

In this chapter, we further assume that both the organization *R* and *S* have a common trusted center (CTC) for determining the group secret parameters of the two groups and the secret shares all members. In any case, of dispute between the group *S* and *R*, the CTC keeps the records of signatures and plays the role of a trusted judge. Since the CTC can checks the validity of the signature so when any third party needs the signature verification, the CTC convince the third party about the facts.

## Chapter 5

***In threshold schemes***, when any *t* or more shareholders act in collusion, they can impersonate any other set of shareholders to forge the signatures. In this case, the malicious set of signers does not have any responsibility for the signatures and it is impossible to trace the signers. Unfortunately, with the ($t, n$) threshold schemes proposed so far, this problem cannot be solved.

***In multisignature schemes,*** the signers of a multisignature are identified in the beginning and the validity of the multisignature has to be verified with the help of identities of the signers. For multisignatures, it is indeed unnecessary to put a threshold value to restrict the number of signers. Consider the situation, where a group of anonymous members would have to generate a multisignature. The members of this group use pseudonyms as their identities in the public directory. What concerns the verifier most is that a message is signed by at least $t$ members and they indeed come from that group. However, the verifier has no way to verify whether a user is in fact a member of that group because of the anonymity of the membership. In this case, the multisignature schemes cannot solve this problems, however, the threshold signature schemes do.

In this chapter, we combine the idea of directed signature scheme and threshold signature with multi signature schemes and propose a digital signature scheme named as *directed - threshold multi - signature scheme*. In this scheme,

- There is a ***trusted SDC*** that determine the group secret keys, all shareholder's secret shares

- There is a ***designated combiner DC*** who takes the responsibility to collect and verify each partial signature and then produce a group signature, ***but no secret information associated with the DC***

- Any malicious set of signers cannot impersonate any other set of signers to forge the signatures. In case of forgery, it is possible to trace the signing set.

- Any *t* or more shareholders act in collusion cannot conspire to reconstruct the polynomial $f_S(x)$ by providing their own secret shares and hence they cannot recover the group secret key.

## Chapter 6

Many threshold signature schemes require a **trusted SDC** to generate the group secret keys and secret shares of group members, which have a single point of vulnerability. The existence of such a center is not a reasonable assumption; there are two potential problems.

- First, for many applications, there is no one person or devices which can be completely trusted by all members of the group.

- Second, the use of a key center creates a single point failure. Any security lapse at the key center can reveal the private key.

To avoid these problems, in 1992, Harn introduced a scheme based on a modified ElGamal signature scheme, ***which does not require a trusted SDC***. In this scheme, each member works as a SDC, generates and distributes the secret shares for each user. In this chapter, we propose a ***directed - threshold multi - signature scheme without SDC***. In this scheme, each shareholder works as a SDC to generate his secret key and distribute the corresponding secret shares to other shareholders. All the shareholders agree in advance for the public parameters. We assume that there is a **designated combiner** *DC* who takes the responsibility to collect and verify each partial signature and then produce a group signature. **Nevertheless, there is no partial secret information of the other users associated with the *DC*.** In this scheme,

- **Each shareholder works as a SDC** to generate his secret key and distribute the corresponding secret shares to other shareholders.

- There is a **designated combiner** *DC* who takes the responsibility to collect and verify each partial signature and then produce a group signature, **but no secret information is associated with the *DC*.**

- Any malicious set of signers cannot impersonate any other set of signers to forge the signatures. In case of forgery, it is possible to trace the signing set.

# Chapter 7

When the message is sensitive to the signature receiver and *requires the approval of some specified subsets of signers*. In this note, we combine the idea of directed signature scheme with threshold multi-signature scheme and propose a threshold digital signature scheme under the title **Generalized Directed - Threshold Multi - Signature Scheme.** The features in this generalized directed- threshold multi - signature scheme are similar to those of the (t ,n) directed threshold- multi – signature scheme as pointed out in the last chapter except that the group signature can be generated only by some specified subsets of members according to the signature policy.

If the message $m$ **(sensitive to the signature receiver $R$)** is transmitted by an organization $S$ to a person $R$ and the message requires the approval of some specific subsets of signers among the set of signers, then the responsibility of signing the messages needs to be shared by some specified subsets of the signers according to the signing policy. For this situation, we propose a **Generalized Directed - threshold multi - signature scheme.**

Let $G_S$ is the set of $n$ signers belonging to the organization $S$ and $H_S$ denote the specified subset of $t$ signers. The valid group signature can be generated by the cooperation of the signers belongs to $H_S$. We also define that $P(G_S)$ is the set of all specified subsets.

For our construction, we assume that there is a **trusted share distribution center ($SDC$) that** determines the group secret keys, all shareholder's secret shares, and a **designated combiner $DC$** who takes the responsibility to collect and verify each partial signature and then produce a group signature. In this scheme,

- There is a $SDC$ to generate group secret key and corresponding secret shares of shareholders.

- There is a **designated combiner $DC$** who takes the responsibility to collect and verify each partial signature and then produce a group signature, **but no secret information is associated with the $DC$.**

- Any malicious set of signers cannot impersonate any other set of signers to forge the signatures. In case of forgery, it is possible to trace the signing set.

- Any $t$ or more shareholders acting in collusion cannot conspire to reconstruct the group secret key by providing their secret shares.

The thesis ends with a list of references (papers, books and websites) that have been consulted from time to time while completing the work.

# Chapter 1
# A Directed Signature Scheme

# A Directed Signature Scheme

## 1.1. Introduction

Already we have discussed about the necessity of the directed signature scheme. Directed signature scheme suits the situation when the message is sensitive to the signature receiver. In this chapter, we propose **a directed signature scheme** and discuss some of its applications. In this scheme, any third party can check the signature validity only with the help of signature receiver.

## 1.2. A Directed Signature Scheme

Suppose that user A wants to generate a signature on the message $m$ so that only receiver B can directed verify the signature and B can prove the validity of signature to any third party C, whenever necessary. The signing and verification processes are as follows.

### 1.2.1. Signature Generation by A

(a). A picks random $K_{a_1}$ and $K_{a_2} \in Z_q$ and computes

$$W_B = g^{-K_{a_2}} \mod p \text{ and } V_B = g^{K_{a_1}} \cdot y_B^{K_{a_2}} \mod p.$$

  Here $y_B$ is the public key of the signature receiver B.

(b). Using a one-way hash function $h$, A computes a secret value $r_A = h(g^{K_{a_1}}, m)$.

(c). A computes $S_A = K_{a_1} + x_A \cdot r_A \mod q$. Here $x_A$ is the private key of the signer.
  $\{S_A, W_B, V_B, m\}$ is the signature of A on the message $m$.

### 1.2.2. Signature Verification by B

(a). Using his private key $x_B$, B computes $R = V_B (W_B)^{x_B} \mod p$ and recovers

$$r_A = h(R, m)$$

(b). B checks the congruence $g^{S_A} \stackrel{?}{\equiv} R \cdot y_A{}^{r_A} \mod p$ for a valid signature. If hold then $\{S_A, W_B, V_B, m\}$ is a valid signature.

### 1.2.3. Proof of Validity by B to C

(a). B picks random $K \in Z_q$ and computes $W_C = g^{-K} \mod p$, $V_C = R \cdot y_C{}^K \mod p$ and sends to C.

(b). C uses $(W_C, V_C)$ in place $(W_B, V_B)$ to checks the validity of signature by using his secret key. The signature verification process will remain same as in sub-section 1.2.2.

## 1.3. Security Discussions

In this sub-section, we shall discuss the security of proposed *Directed Signature Scheme.*

(a). Can one retrieve the secret key $x_A$, integer $K_{a_1}$ and secret value $r_A$ from the equation $S_A = K_{a_1} + x_A \cdot r_A \mod q$?

Here the numbers of unknown parameters are three. The number of equation is one, so it is computationally infeasible for a forger to collect the secret key $x_A$, integer $K_{a_1}$ and secret value $r_A$ from this equation.

(b). Can one impersonate the signer ?

A forger may try to impersonate the signer by randomly selecting two integers $K_1, K_2 \in Z_q$ and calculate

$W_B = g^{-K_2} \mod p$, $V_B = g^{K_1} y_B{}^{K_2} \mod p$ and $r_A = h(g^{K_1}, m)$.

But without knowing the secret key $x_A$, it is difficult to generate a valid $S_A$ to satisfy the verification equation, $g^{S_A} \stackrel{?}{\equiv} [R \cdot y_A{}^{r_A}] \mod p$.

(c). Can one forge a signature $\{S_A, W_B, V_B, m\}$ by the equation,

$$g^{S_A} \stackrel{?}{\equiv} [R \cdot y_A{}^{r_A}] \mod p?$$

A forger may randomly select an integer R and then computes the hash value $r_A$ such that $r_A = h(R, m) \mod q$. Obviously, to compute the integer $S_A$ is equivalent to solving the discrete logarithm problem. Thus these attacks will not be successful.

## Illustration

To illustrate the scheme, taking $p = 23$, $q = 11$ and $g = 3$. The secret and public key of users is as follow.

| User | Secret key | public key |
|------|------------|------------|
| A | 4 | 12 |
| B | 7 | 2 |
| C | 6 | 16 |

To generate the signature, A picks random $K_{a_1} = 9$, and $K_{a_2} = 5$ and computes $W_B = 16, V_B = 1$. Using a one way hash function $h$, A computes $r_A = 10$, $S_A = 5$ sends $\{5, 16, 1, m\}$ to B as his/her signature on the message $m$.

B computes R = 18, recovers $r_A = h(18, m) = 10$ (Let). B checks the congruence $3^5 \stackrel{?}{\equiv} 18 \cdot 12^{10} \mod 23$ for a valid signature. This holds.

To prove the validity of the signature to C, the signature receiver B picks random $K = 8$ and computes $W_C = 4$, $V_C = 9$ and sends to C. C computes $R = 9 \cdot 4^6 \mod 23 = 18$. C verifies the signature in the similar fashion as signature receiver.

## 1.4. A Directed Signature with Threshold Verification

Our directed signature scheme can be used to design a signature scheme with threshold verification. A signer can generate a signature on the message $m$ to the organization $R$ in such a way that any subset $H_R$ of $k$ or more users of a designated group $G_R$ of $n$ users can determine the validity of the signature but any $k-1$ or less users cannot verify the signature. The construction of *a directed signature scheme with threshold verification* is based on Shamir's threshold scheme [97]. We assume that every user $i$ in the designated group $G_R$ possesses a private key $x_{R_i}$ and public key $y_{R_i}$.

This scheme consists of the following steps.

## 1.4.1. Signature Generation by A

(a). A picks random $K_{a_1}$ and $K_{a_2} \in Z_q$ and computes

$$W_R = g^{-K_{a_2}} \bmod p \text{ and } V_R = g^{K_{a_1}} \bmod p.$$

(b). Using a one-way hash function $h$, A computes a secret value

$$r_A = h(V_R, m) \text{ and } S_A = K_{a_1} + x_A \cdot r_A \bmod q.$$

Here $x_A$ is the private key of the signer.

(c). A selects a polynomial $f_R(x) = K_{a_1} + b_1 x + \ldots + b_{k-1} x^{k-1} \bmod q$, with $K_{a_1} = f_R(0)$.

(d). A computes a public value $v_{R_i}$ for each member of the group $G_R$, as,

$$v_{R_i} = f_R(u_{R_i}) \cdot y_{R_i}^{K_{a_2}} \bmod p.$$

Here $y_{R_i}$ is public key and $u_{R_i}$ is the public value associated with each user $i$ in the group $G_R$.

(e). A sends $\{S_A, W_R, m, \{v_{R_i}\}_{i=1}^n\}$ to the group $G_R$ as the signature of A.

## 1.4.2. Signature Verification by the Organization R

Any subset $H_R$ of $k$ users from a designated group $G_R$ of $n$ users can verify the signature. We assume that there is a designated combiner, which collects partial computations from each user in $H_R$ and determine the validity of signature. The verifying process is as follows.

(a). Each user $i \in H_R$ recovers his/her secret share $f_R(u_{R_i}) = v_{R_i} W_R^{x_{R_i}} \bmod p$.

(b). Each user $i \in H_R$ modifies his/her shadow, as,

$$MS_{R_i} = f_R(u_{R_i}) \cdot \prod_{j=1, j \neq i}^{t} \frac{-u_{R_j}}{u_{R_i} - u_{R_j}} \mod q.$$

(c) Each user $i \in H_R$ uses his/her modified shadow, $MS_{R_i}$, and calculate the partial result $R_{R_i} = g^{MS_{R_i}} \mod q$ and sends to the combiner.

(d). The combiner computes $R = \prod_{i=1}^{k} R_{R_i} \mod p$ and recovers $r_A = h(R, m) \mod q$.

(e). The combiner checks the congruence $g^{S_A} \stackrel{?}{\equiv} R \cdot y_A^{r_A} \mod p$ for a valid signature.

If hold then $\{\{S_A, W_R, m, \{v_{R_i}\}_{i=1}^{n}\}\}$ is a valid signature of the signer A.

The proposed scheme has the following characteristics:

- There is no need of a trusted SDC, while in many threshold signature schemes, there is a trusted SDC that determines the security parameters and also distributes the secrets shares to each user in the system.

- The signer is free to decide the security parameters and secrets shares to each user.

- The signer decides the security parameters and secrets shares at the time of signing.

- The signer can use the security parameters and secrets shares as one-time secrets. There is no need to fix them.

- The signer can change the threshold value $k$ to sign another documents.

- The only requirement is that every user $i$ in the designated group $G_R$ possesses a private and public key pair $(x_{R_i}, y_{R_i})$.

## 1.5. Application to Threshold Cryptosystem

Threshold cryptosystem is society – oriented cryptosystem. This is useful, when a policy requires the consent of more than one person. The organization has a single public key. The secret key is distributed among the users of organization in such a way that any $k$ or more users can determines the policy, but when any $k-1$ or less users works together, they cannot receive any information about the system secret.

Desmedt. Y. and Frankel. Y. [20,21,22,23] developed the concepts of threshold cryptosystems. They're some weaknesses in these cryptosystem, given as follows:

- The group secret key is fixed; any subgroup of $k$ dishonest users can recovers the group secret key and can be harmful for the system.

- The security parameters are also fixed. The secrets shares determine during the system set up. Any change in-group member or security policy requires adjusting group member's secret share accordingly.

Our directed signature scheme can be used to make a threshold cryptosystem to sort out all the above weaknesses. Suppose a user A wants to encrypt the message $m$ so that any $k$ users from designated group $G_R$ of $n$ users should cooperate to recover the message. The encryption and decryption key $K$ for sender and the receiving group $G_R$ can be recovered from $V_R$ and $R$ respectively. The sender A broadcasts $\{\{S_A, W_R, c, \{v_{R_i}\}_{i=1}^{n}\}$ to the group $G_R$, where $c = E_K(m)$ and $K = h(V_R)$. Any $k$ members can recover $R$ and the decryption key $K = h(R)$. Therefore they can decrypt the message by $D_K(c) = m$.

The proposed threshold cryptosystem has the following advantages over the cryptosystems based on the concept of Desmedt and Frankel.

- The group secret is not fixed and can be change for each further communications.

- There is no need of fixed set up. The sender determines everything at the time of encryption.

- An individual can send cipher text to any chosen group with any desired security policy.

# Chapter 2
# A Directed Delegated Signature Scheme

# A Directed-Delegated Signature Scheme

## 2.1 Introduction

Delegation of rights is a common practice in the real world. In the electronic world to facilitate this requirement, proxy signature schemes have been invented through which signing capability are delegate efficiently and transparently. Digital signature schemes rely on a secret signature key [4], which only the certified person knows. If this secret key is delegated to another person directly, it can no longer be identified with that person and hence the assumption of the digital signature is broken. Therefore, there is need to delegate the signing capability without revealing the secret key such that the recipient can verify the signature of the original signer with the help of the delegated or proxy signer.

On the other hand, in the situation, when the message is sensitive to the signature receiver directed signature are used. Consider a different situation, when we want to delegate the signing capability and the message is sensitive to the signature receiver. For this situation, we combine the idea of proxy signature scheme with directed signature scheme and obtains a ***Directed-delegated signature scheme.***

In this chapter we propose a directed – delegated signature scheme, in which the signing of message is done in two phases. The first phase is off-line. This phase is governed by proxy signature scheme, and can be performed even before the message to be signed is known. The second phase is on-line. It starts after the message becomes known and utilizes the precomputation of the first phase.

In this chapter, we also illustrate an application of the proposed ***directed-delegated signature scheme***. consider a central authority A, who designates a trusted authority B to issue certificate on his behalf. The message is sensitive for C and is also of concern to other users. C wants a certificate that nobody can check the validity of signature without his help, but C can prove the validity of the signature to any third party Y, whenever necessary.

## 2.2. A Directed – Delegated Signature Scheme

Consider a situation, where, there is a NGO's hospital which facilitates AIDS checkup. The hospital is headed by a CMO A, who has many responsibility to perform. He designates a trusted colleague B to issue certificate on his behalf. A patient C does not want to disclose the result of his/her check-up and he/she wants a certificate that nobody can read without his help. Unfortunately if C is HIV positive then there is need of curing. For treatment there is another important hospital headed by a NGO chief Y, Where all types of Medicine, remedies and all resources are available for AIDS patients C. For treatment there, C has to prove the validity of a certificate to a desired person .

The one solution of such problems is governed by directed - delegated signature schemes. Here the receiver C can directly verify the signature and that C can prove its validity to any third party Y, whenever necessary. Since directed signature can be verified with the help of receiver C only so we can say that the contents of certificates have no validity without the signature verification. The following is an exposition on how proxy and directed signature scheme can be implemented for our construction.

Before organizing the check up, A appoints B as a designated signer. He delivers a designated signature key to B using the following protocol. A chooses a large prime number $p$, $q$ a large prime factor of $p-1$, an integer $g \in Z_p$ with order $q$, a one way hash function $h$.

### 2.2.1. Signature Key Delegation by A

1. A selects $k_A \in Z_q$, computes $r_A = g^{k_A} \mod p$ and sends $r_A$ to B.

2. (a). B randomly selects $\alpha \in Z_q$ and computes $r = g^{\alpha} r_A \mod p$.

   (b) If $r \in Z_q^*$, he sends $r$ to A, otherwise goes to step (a).

3. A computes $s_A = r x_A + k_A \mod q$ and forwards $s_A$ to B.

4. B computes $S = s_A + \alpha \mod q$ and check $g^S = y^r r \mod p$.

If equality holds, B accepts '$s_A$' as a valid designated signature key from A.

Now the hospital is open for the public. B does the check up of a patient C and prepares a report *m.* B gives a certified report to C. The signature generation and verification is done by the following protocol.

## 2.2.2. Signature Generation by B for C

(a). B picks at random $K_{b_1}$ and $K_{b_2} \in Z_q$ and computes

$$W_B = g^{K_{b_1} - K_{b_2}} \bmod p \text{ and } Z_C = y_C^{K_{b_1}} \bmod p.$$

(b). B computes $r_B = h(Z_C, W_B, m)$ and $S_B = K_{b_2} - S \cdot r_B \bmod q$.

(c). B sends $\{S_B, W_B, r_B, m\}$ to C as CMO's signature.

## 2.2..3. Signature Verification by C

(a). C computes $\mu = (g^{S_B}(y_A^{r}.r)^{r_B} W_B) \bmod p$ and $Z_C = \mu^{x_C} \bmod p$.

(b). C checks that $r_B = h(Z_C, W_B, m)$ for the validity of signature.

## 2.2.4. Proof of Validity by C to Y

(a) C sends to $\{S_A, W_B, r_A, m, \mu\}$ to Y.

(b) Y checks if $r_A = h(Z_B, W_B, m) \bmod q$.

If this does not hold Y stops the process; otherwise goes to the next steps.

(c) C in a zero knowledge fashion proves to Y that $\log_\mu Z_C = \log_g y_C$ as follows.

- Y chooses random $u, v \in Z_p$ computes $w = \mu^u \cdot g^v \bmod p$ and sends $w$ to C.

- C chooses random $\alpha \in Z_p$ computes $\beta = w \cdot g^\alpha \bmod p$ and $\gamma = \beta^{x_B} \bmod p$ and sends to Y.

- Y sends $u, v$ to B, by which C can verify that $w = \mu^u \cdot g^v \bmod p$.

- C sends $\alpha$ to Y, by which she can verify that

$$\beta = \mu^u \cdot g^{v+\alpha} \bmod p \text{ and } \gamma = Z_C^u \cdot y_C^{v+\alpha} \bmod p.$$

## 2.3. Security Discussions

In this section, we discuss some possible attacks.

(a). If the designated proxy signer B is dishonest then he can cheat the original signer A and get her/his signature ($r$, $s_A$) on any message $m$ of her/his choice.

The solution of this problem is the existence of a trusted third party. The original signer A may stress that all messages between two parties A and B during the key delegation protocol be authenticated. The third party keeps the records of original signer's orders, and check any case of designated signer disobeying original signer's order.

(b). Can one get integer $K_{b_2}$ and $S$ (secret signature key of proxy signer), from the equation $S_B = K_{b_2} - S \cdot r_B \mod q$?

Here the number of unknown parameters are two. The number of equation is one, so it is computationally infeasible for a forger to collect the secret integer $K_{b_2}$ and $S$.

(c). Can one impersonate the designated signer B?

A forger may try to impersonate the designated signer B by randomly selecting two integers $K_{i_1}$ and $K_{i_2} \in Z_q$. But without knowing the secret part $\alpha$, it is difficult to generate a valid proxy signature key $S$ and $S_B$ to satisfy the verification equation,

$$Z_c = (g^{S_B} (y_A{}^r \cdot r)^{r_B} W_B)^{x_C} \mod p, \quad r_B = h(Z_c, W_B, m).$$

(d). Can one forge a signature { $S_B$, $W_B$, $r_B$, $m$, $r$} using the equation

$$\mu = (g^{S_B} (y_A{}^r \cdot r)^{r_B} W_B) \mod p?$$

To compute the integer $S_B$ from this equation, is equivalent to solving the discrete logarithm problem. If any forger randomly selects $S^*$ and sends { $S^*$, $W_B$, $r_B$, $m$, $r$} to B, the receiver B computes $\mu^* = [g^{S^*}(y_A{}^r \cdot r)^{r_B} W_B]$ mod $p$, $Z^* = \mu^{*x_B} \mod p$ and can check if

$$r_B = h(Z^*, W_B, m), \text{ to detect the forgery.}$$

## Illustration

We illustrate this scheme using small parameters. Taking $p = 23$, $q = 11$, $g = 6$ and the secret keys and the public keys of users are as follows.

| Users | Secret key | public key |
|-------|------------|------------|
| A | 3 | 9 |
| B | 5 | 2 |
| C | 6 | 12 |
| Y | 8 | 18 |

To deliver the signing key, A selects $k_A = 7$ and sends $r_A = 3$ to B. B selects $\alpha = 5$ and sends $r = 6$ to A. A computes $s_A = 3$ and sends to B. B computes proxy signature key $s = 8$ and checks if $6^8 = [(9^6 \cdot 6)] \mod 23$.

To generate the signature, B picks at random $K_{b_1} = 7$, $K_{b_2} = 2$ and calculate $W_B = 2$ and $Z_C = 16$, $r_B = h(16, 2, 0, 8, 3, 18) = 2$ (let) $S_B = 8$, and sends the cryptogram $\{2, 2, 8, 0, 8, 3, 18, 6\}$ to C.

For the signature verification, C computes $\mu = 3$ and $Z_C = 16$ and checks $r_B = h(16, 2, 0, 8, 3, 18) = 2$, for the validity of signature.

To Proof of validity by C to Y, C sends $(16, 2, 2, 8, 3, 0, 8, 3, 18, 3)$ to Y. Y checks $r_B = h(16, 2, 0, 8, 3, 18) = 2$. This holds. Now C can prove that $\log_3 16 = \log_6 12$, in a zero knowledge fashion by using the following confirmation protocol.

(i). Y chooses at random $u = 13, v = 15$ and computes $w = 3$ and sends $w$ to C.

(ii). C chooses at random $a = 8$ and computes $\beta = 8, \gamma = 13$ and sends $\beta, \gamma$ to Y.

(iii). Y sends $u, v$ to C, by which C can verify that $w = 3$.

(iv). C sends $a$ to Y, by which she can verify that $\beta = 16$ and $\gamma = 4$.

## 2.4. Remarks

In this chapter, we have discussed a directed - delegated signature scheme, which is useful in that case when the signed message is sensitive to signature receiver. In this scheme, the signature receiver C has full control over the signature verification process. Nobody can check the validity of signature without his co-operation. The receiver is also able to prove the validity of the signature, whenever necessary.

We have presented a construction of such a directed signature scheme, which is based on discrete logarithm problems. Hence the security level of this scheme is similar to that of other scheme based on discrete logarithm. Since the relation between the signer and signer's secret key is not known to anyone, this scheme is more secure than any other scheme, based on the discrete logarithm.

# Chapter 3

# Directed – Threshold Signature Scheme

# Directed – Threshold Signature Scheme

## 3.1. Introduction

In many situations, the message is on behalf of an organization and a valid message may require the approval or consent of several people. In this case, the signature is done by more than one consenting rather than by a single person. A common example of this policy is a large bank transaction, which requires the signature of more than one person. Such a policy could be implemented by having a separate digital signature for every required signer, but this solution increases the effort to verify the message linearly with the number of signer. **Threshold signature** is an answer to this problem. The (t, n) threshold signature schemes are used to solve these problems. Threshold signatures are closely related to the concept of threshold cryptography, first introduced by Desmedt [21,22,23]. In 1991 Desmedt and Frankel [22] proposed the first (t, n) threshold digital signature scheme based on the RSA assumption.

In this chapter, we propose a **(t , n) directed - threshold signature scheme** based on Shamir's threshold signature scheme[98] and Schnorr's signature scheme[94].

## 3.2. Directed -Threshold Signature Scheme

Let $G$ be a group of $n$ designated users, out of which any $t$ or more members can generate the signature on a message $m$ for a user B. The user B can verify the signature and that B can prove its validity to any third party C, whenever necessary. Nobody can check the validity of the signature without the help of B.

For our construction, we assume the existence of a trusted share distribution center (SDC) which determines the group secrets parameters and the secret shares $v_i$ , $i \in G$. Let $H$ be any subset of G, containing $t$ members. We also assume the existence of a designated combiner $DC$ who collects partial signatures from each user in the subgroup $H$. Every shareholder in the group has equal authority with regard to the group secret. The generation of the directed signature requires $t$ out of $n$ shareholders and interaction with $DC$.

This scheme consists of the following steps.

## 3.2.1. Group Secret Key and Secret Shares Generation

(a). SDC selects the group public parameters $p, q, g$ and a collision free one way hash function $h$. SDC also selects a polynomial

$$f(x) = a_0 + a_1 x + \ldots a_{t-1} x^{t-1} \bmod q, \text{ with } a_0 = x_G = f(0).$$

Here $x_G$ is the secret key of the group $G$.

(b). SDC computes the group public key, $y_G$, as, $y_G = g^{f(0)} \bmod p$.

(c) SDC computes a secret shares $v_i$ for each member of the group $G$, as,

$$v_i = f(u_i) \bmod q,$$

Here $u_i$ is the public value associated with user $i$ in the group $G$.

(d) SDC sends $v_i$ to each user in a secret manner.

## 3.2.2. Signature Generation by any $t$ Users

If any $t$ out of $n$ members of the group agree to sign a message $m$ for user B, they generate the signature using following steps:

(a). Each member $i$ randomly selects $K_{i_1}$ and $K_{i_2} \in Z_q$ and computes

$$w_i = g^{K_{i_2} - K_{i_1}} \bmod p \text{ and } z_i = y_B^{K_{i_2}} \bmod p.$$

(b). Each member makes $w_i$ openly and $z_i$ secretly available to each member of $H$. Once all $w_i$ and $z_i$ are available, every member computes $Z, W$ and $R$ as

$$W = \prod_{i \in H} w_i \bmod q, \quad Z = \prod_{i \in H} z_i \bmod q \text{ and } R = h(Z, W, m) \bmod q.$$

(c). Each member $i$ modifies his/her share, as $MS_i = v_i \cdot \prod_{j=1, j \neq i}^{t} \frac{-u_j}{u_i - u_j} \bmod q$.

(d). Each member $i$ uses his/her modified share, $MS_i$ and random integer $K_{i_1}$ to calculate the partial signature $s_i$ as, $s_i = K_{i_1} - MS_i.R \bmod q$.

(e). Each member $i$ sends his/her partial signature to the designated combiner $DC$ who collects the partial signatures and produces the group signature

$$S = \sum_{i=1,}^{t} s_i \bmod q.$$

(f) DC sends $\{S, W, R, m\}$ to B as signature of the group $G$ on the message $m$.

### 3.2.3. Signature Verification by B

(a). B computes $\mu = [\, g^S \, (y_G)^R \, W\,] \bmod p$ and $Z = \mu^{x_B} \bmod p$.

(b) B checks the validity of signature by verifying $R = h(Z, W, m) \bmod q$.

### 3.2.4. Proof of Validity By B to Any Third Party C

This part of the protocol runs as in section- 2.2.4.

## 3.3. Security Discussions

we now discuss security aspect of the proposed **Directed - Threshold Signature Scheme**.

(a). Is it possible to retrieve the group secret key $f(0)$ from the group public key $y_G$? No because this is as difficult as solving discrete logarithm problem.

(b). Can one retrieve the secret shares $v_i$, $i \in G$, from the public value $u_i$? No because $f$ is a randomly and secretly selected polynomial.

(c). Can one retrieve the secret shares $v_i$, integer $K_{i_1}$ and partial signature $s_i$, from the equation $s_i = K_{i_1} - MS_i.R \bmod q$. ?

Here the number of unknown parameters is two. The number of equation is one, so it is computationally infeasible for a forger to collect the secret shares $v_i$, integer $K_{i_1}$ and partial signature $s_i$, $i \in G$.

**(d).** Can the designated combiner $DC$ retrieve the group secret key $f(0)$ or any partial information from the equation, $S = \sum_{i=1}^{t} s_i \bmod q$ ?

This is again computationally infeasible.

**(e).** Can one impersonate a member $i, i \in H$ ?

A forger may try to impersonate a shareholder $i$, by randomly selecting two integers $K_{i_1}$ and $K_{i_2} \in Zq$ and broadcasting $w_i$ and $z_i$. But without knowing the secret share $v_i$, it is difficult to generate a valid partial signature $s_i$ to satisfy the verification equation,

$$Z = [\, g^S (y_G)^R \; W ]^{x_B} \bmod p, \text{ where }, S = \sum_{i=1}^{t} s_i \bmod q.$$

**(f).** Can one forge a signature $\{S, W, R, m\}$ by the following equation,

$$\mu = [\, g^S (y_G)^R \; W ] \bmod p.?$$

To compute integer $S$ from here is equivalent to solving the discrete logarithm problem. If any forger randomly selects $S^*$ and sends $\{S^*, W, Z, R, m\}$ to B, the receiver B would computes

$$\mu^* = [\, g^{S^*} (y_G)^R \; W ] \bmod p, \text{ and } Z = \mu^{*x_B} \bmod p \text{ and checks if}$$

$$r_B \stackrel{?}{=} h(Z^*, W_B, m).$$

The receiver B can easily identify that someone has forged the signature.

**(g).** Can $t$ or more shareholders act in collusion to reconstruct the polynomial $f(x)$ .?

According to the following equation,

$$f(x) = \sum_{i=1}^{t} f(u_i) \prod_{j=1, j \neq i}^{t} \frac{x - u_j}{u_i - u_j} \bmod q,$$

the secret polynomial $f(x)$ can be reconstructed with the knowledge of any $t$ secret shares $f(u_i)$, $i \in G$.

So if in an organization the shareholders are known to each other, the temptation for $t$ of them to collude could be irresistible. As a result, they would find the secret key of the company, which will be continued to be used.

This attack does not weaken the security of our scheme in the sense that the number of users that have to collude in order to forge the signature is not smaller than the threshold value. However it is worth pointing out that $t$ or more users can conspire to compute the system secrets.

## Illustration

For illustration, we assume there are four users in the system. Out of four users A, C, E and F any two users, say, A and F can generate the directed signature on a message $m$ for the user B with secret and public key pair $x_B = 6$, $y_B = 8$ respectively. The following steps illustrate this scheme.

### Group Secret Key and Secret Shares Generation

Let SDC choose $p = 23$, $q = 11$, $g = 18$ and $f(x) = 3 + 5x \bmod 11$, where $f(0) = 3$, is the group secret key. The public values $u_i$ and corresponding secret shares $v_i$ of users are as follow----

| Users | public value ($u_i$) | secret share ($v_i$) |
|---|---|---|
| A | 9 | 4 |
| C | 12 | 8 |
| E | 14 | 7 |
| F | 16 | 6 |

Now the SDC determines the group secret key as $f(0)$ and computes the group public key, $y_G$, as $y_G = 18^3 \bmod 23 = 13$.

### Signature Generation by any $t$ Users

If any two users A and F out of four users agree to sign a message $m$ for user B, then the signature generation has the following steps.

(a) The user A randomly selects $K_{a_1} = 2$, $K_{a_2} = 7$ and computes $w_1 = 3$, $z_1 = 12$.

(b) The user F randomly selects $K_{f_1} = 5$, $K_{f_2} = 9$ and computes $w_4 = 4$, $z_4 = 9$.

(c) Both the users A and F make $(w_1, w_4)$ and $(z_1, z_4)$ publicly available through a broadcast channel. Once all $(w_1, w_4)$ and $(z_1, z_4)$ are available, each user in $H$ computes the product $W = 12$, $Z = 16$ and $R = h(16, 12, m)$ mod $11 = 5$ (Let us suppose that, this value is turn out equal to 5).

(c) The users A and F compute their modified shares $MS_A = 6$ and $MS_F = 8$.

(d) The users A and the user F calculates his/her partial signature $s_1 = 5$ and $s_2 = 9$ respectively. Both the signers send their partial signature to $DC$ who produces a group signature $S = 3$.

(g) DC sends $\{3, 12, 5, m\}$ to B as signature of the group $G$ for the message $m$.

### Signature Verification by B

B computes $\mu = [18^3 \cdot 13^{5} \cdot 12]$ mod $23 = 3$ and $Z = 16$ and checks the validity of signature by computing $R = h(16, 12, m)$ mod $11 = 5$.

### Proof of Validity by B to any Third Party C

B sends $\{3, 12, 5, m, 3\}$ to C, C checks that $R = h(16, 12, m)$ mod $11 = 5$. This holds. Now B proves to C that $\log_3 16 = \log_{18} 8$ in a zero knowledge fashion by using the following confirmation protocol.

(i). C chooses at random $u = 11$, $v = 13$ and computes $w = 2$ and sends $w$ to B.

(ii). B chooses at random $α = 17$ and computes $ß = 16$, $γ = 4$ and sends $ß$, $γ$ to C.

(iii). C sends $u$, $v$ to B, by which B can verify that $w = 2$.

(iv). B sends $α$ to C, by which she can verify that $ß = 16$ and $γ = 4$

## 3.4. Remarks

In this scheme, we have used the ElGamal public key cryptosystem to obtain the construction of Directed Threshold Signature Scheme. The security of this cryptosystem is based on the discrete log problem. Only $t - 1$ shadows are not sufficient to obtain the group secret key and they will also get no information about the group secret key, until $t$ individuals act in collusion.

In this scheme, there is a **DC** who collects the partial signature of the signer. We should note that there is no secret information associated with the *DC*. Every user can compute his modified share under mod $q$. If $q$ is not prime, then the calculations of the exponents is performed mod $\Phi(q)$, which is not a prime. This implies that Lagrange interpolation for calculating the modified shadows will not work (except when $q = 3$, in which case we are not interested). Consider the situation, when $\prod_{j=1, j \neq i}^{t}(u_i - u_j)$ and $q$ are coprime. In this case there is no way to find out the multiplicative inverse of $\prod_{j=1, j \neq i}^{t}(u_i - u_j) \bmod q$. There is only possibility of selecting the large prime $q$ numbers in order for each person to get around this difficulty

Directed threshold signatures are meaningless to any third party because there is no way for him to prove its validity. The only knowledge of $Z$ is not sufficient to prove the validity of signature. Signature receiver also has to perform the confirmation protocol in a zero knowledge fashion to prove the validity of signature.

# Chapter 4
# Threshold Signature Scheme with Threshold Verification

# Threshold Signature Scheme with Threshold Verification

## 4.1. Introduction

In most situations, the signer and the verifier is generally a single person. However, when the message is sent by one organization to another organization, a valid message may require the approval or consent of several people. In this case, the signature generation and verification is done by more than one consenting rather than by a single person. This chapter proposes a *Threshold signature scheme with threshold verification.*

## 4.2. Threshold Signature Scheme with Threshold Verification

Consider the message is transmitted by an organization $S$ to another organization $R$ and may require the approval of a subset $H_S$ of $t$ or more signer from a designated group $G_S$ of $n$ users belongs to the organization $S$.

On the other hand, the signing group wants to generate the signature on a message $m$ in such a way that the signature can be verified by any subset $H_R$ of $k$ or more users from a designated group $G_R$ of $l$ users belongs to the organization $R$, then threshold verification schemes serve our purpose.

We assume that the secret key of the organization $S$ is $x_S$ and the public key is $y_S$ where $y_S = g^{x_S} \mod p$ with $x_S \in Z_q$. Similarly the organization $R$ possesses a pair $(x_R, y_R)$ of private key $x_R$ and public key $y_R = g^{x_R} \mod p$. In addition, every user A in both the organizations possesses a pair $(x_A, y_A)$ with $x_A$ secret and $y_A = g^{x_A} \mod p$ public.

We further assume that both the organization $R$ and $S$ have a common trusted center (CTC) for determining the group secret parameters of the two groups and the secret shares all members. This scheme consists of the following steps.

## 4.2.1. Group Secret Key and Secret Shares Generation for the Organization $S$

(a). CTC selects the group public parameters $p$, $q$, $g$ and a collision free one way hash function $h$. CTC also selects a polynomial $f_S(x)$ for the group $G_S$, as,

$$f_S(x) = a_0 + a_1 x + \ldots a_{t-1} x^{t-1} \bmod q, \text{ with } a_0 = x_S = f_S(0).$$

(b). CTC computes the group public key, $y_S = g^{f_S(0)} \bmod p$.

(c). CTC randomly selects $K \in Z_q$ and computes a public value $W = g^{-K} \bmod p$.

(d). CTC computes a public value $v_{S_i}$ for each member of the group $G_S$, as,

$$v_{S_i} = f_S(u_{S_i}) \cdot y_{S_i}^{K} \bmod p.$$

Here, $y_{S_i}$ is public key and $u_{S_i}$ is the public value associated with each user $i$ in the group $G_S$.

(e). CTC sends $\{v_{S_i}, W\}$ to each user $i$ in the group $G_S$ through a public channel.

## 4.2.2. Group Secret Key and Secret Shares Generation for the Organization $R$

(a). CTC selects a polynomial $f_R(x)$ for the group $G_R$, as,

$$f_R(x) = b_0 + b_1 x + \ldots b_{k-1} x^{k-1} \bmod q \text{ with } b_0 = x_R = f_R(0).$$

(b). CTC computes the group public key, $y_R = g^{f_R(0)} \bmod p$.

(c). CTC computes a public value $v_{R_i}$ for each member of the group $G_R$, as,

$$v_{R_i} = f_R(u_{R_i}) \cdot y_{R_i}^{K} \bmod p.$$

Here, $y_{R_i}$ is public key and $u_{R_i}$ is the public value associated with each user $i$ in the group $G_R$.

(d). CTC sends $\{v_{R_i}, W\}$ to each user $i$ in the group $G_R$ through a public channel.

### 4.2.3. Signature Generation by any *t* Users

If $H_S$ is a subset of $t$ members of the organization $S$ out of $n$ members who agree to sign a message $m$ to be sent to the organization $R$, then the signature generation has the following steps:

(a). Each user $i \in H_S$ randomly selects $K_{i_1}$ and $K_{i_2} \in Z_q$ and computes

$$u_i = g^{-K_{i_2}} \bmod p, \quad v_i = g^{K_{i_1}} \bmod p \text{ and } w_i = g^{K_{i_1}} y_R^{K_{i_2}} \bmod p.$$

(b). Each user $i$ broadcasts $u_i$, $w_i$ publicly and $v_i$ secretly to every other user in $H_S$. Once all $u_i, v_i$ and $w_i$ are available, each member $i, i \in H_S$ computes the product $U_S$, $V_S$, $W_S$ and a hash value $R_S$, as,

$$U_S = \prod_{i \in H_S} u_i \bmod q, \quad V_S = \prod_{i \in H_S} v_i \bmod q,$$

$$W_S = \prod_{i \in H_S} w_i \bmod q \text{ and } R_S = h(V_S, m) \bmod q.$$

(c). Each user $i \in H_S$ recovers his/her secret share $f_S(u_{S_i}) = v_{S_i} W^{x_{S_i}} \bmod p$.

(d). Each user $i \in H_S$ modifies his/her shadow $MS_{S_i} = f_S(u_{S_i}) \cdot \prod_{j=1, j \neq i}^{t} \dfrac{-u_{S_j}}{u_{S_i} - u_{S_j}} \bmod q$.

(e). By using his/her modified shadow $MS_{S_i}$, each user $i \in H_S$ computes his/her partial signature $s_i = K_{i_1} + MS_{S_i} \cdot R_S \bmod q$.

(f). Each user $i \in H_S$ sends his/her partial signature to the CTC, who produce a group signature $S_S = \sum_{i=1}^{t} s_i \bmod q$.

(g). CTC sends $\{S_S, U_S, W_S, m\}$ to the **designated combiner *DC* of organization R** as signature of the group *S* for the message *m*.

## 4.2.4. Signature Verification by the Organization *R*

Any subset $H_R$ of *k* users from a designated group $G_R$ can verify the signature. We assume that there is **designated combiner *DC* (can be any member among the members of the group $G_R$, or the head of the organization *R*)**, who collects partial computations from each user in $H_R$ and determines the validity of signature. The verifying process is as follows.

(a). Each user $i \in H_R$ recovers his/her secret share $f_R(u_{R_i}) = v_{R_i} W^{x_{R_i}} \mod p$.

(b). Each user $i \in H_R$ modifies his/her shadow $MS_{R_i} = f_R(u_{R_i}) \prod_{j=1, j \neq i}^{k} \frac{-u_{R_i}}{u_{R_i} - u_{R_j}} \mod q$.

(c) Each user $i \in H_R$ sends his modified shadow $MS_{R_i}$ to the *DC*.

(d). *DC* computes $R_R$, as, $R_R = W_S . U_S^{\sum_{i=1}^{k} MS_{R_i}} \mod q$ and recovers

$$R_S = h(R_R, m) \mod q.$$

(d). *DC* checks the congruence $g^{S_S} \stackrel{?}{\equiv} R_R . y_S^{R_S} \mod p$ for a valid signature. If hold then $\{S_S, U_S, W_S, m\}$ is a valid signature on the message *m*.

## 4.3. Security Discussions

In this sub-section, we shall discuss the security aspects of proposed *Threshold Signature Scheme with threshold verification.* Here we shall discuss several possible attacks. But none of these can successfully break our system.

**(a).** Can any one retrieve the organization's secret keys $x_S$ and $x_R$ from the group public key $y_S$ and $y_R$ respectively ?

This is as difficult as solving discrete logarithm problem. No one can get the secret key $x_S$ and $x_R$, since $f_S$ and $f_R$ are the randomly and secretly selected polynomials by the CTC. On the other hand, by using the public keys $y_S$ and $y_R$ no one also get the secret keys $x_S$ and $x_R$ because this is as difficult as solving discrete logarithm problem.

(b). Can one retrieve the secret shares, $f_S(u_{S_i})$ of members of $G_S$, from the equation $v_{S_i} = f_S(u_{S_i}) \cdot y_{S_i}^K \mod p$ ?

No because $f_S$ is a randomly and secretly selected polynomial and $K$ is also a randomly and secretly selected integer by the CTC. Similarly no one can get the secret shares $f_R(u_{R_i})$ of members of $G_R$, from the equation,

$$v_{R_i} = f_R(u_{R_i}) \cdot y_{R_i}^K \mod p.$$

(c). Can one retrieve the secret shares, $f_S(u_{S_i})$ of members of $G_S$, from the equation $f_S(u_{S_i}) = v_{S_i} W^{x_{S_i}} \mod p$ ?

Only the user $i$ can recovers his secret shares, $f_S(u_{S_i})$, because $f_S$ is a randomly and secretly selected polynomial and $x_{S_i}$ is secret key of the user $i \in G_S$. Similarly no one can get the secret shares $f_R(u_{R_i})$ of members of $G_R$, from the equation,

$$f_R(u_{R_i}) = v_{R_i} W^{x_{R_i}} \mod p.$$

(d). Can one retrieve the modified shadow $MS_{S_i}$, integer $K_{i_1}$, the hash value $R_S$ and partial signature $s_i, i \in G_S$ from the equation

$$s_i = K_{i_1} + MS_{S_i} \cdot R_S \mod q ?$$

They all are secret parameters and it is computationally infeasible for a forger to collect the $MS_{S_i}$, integer $K_{i_1}$, the hash value $R_S$ and partial signature $s_i, i \in G_S$.

(e). Can the designated CTC retrieve any partial information from the equation,

$$S_S = \sum_{i=1,}^{t} s_i \mod q ?$$

Obviously, it would be computationally infeasible for CTC to derive any information from $s_i$.

(f). Can one impersonate a member $i, i \in H_S$ ?

A forger may try to impersonate a shareholder $i, i \in H_S$, by randomly selecting two integers $K_{i_1}$ and $K_{i_2} \in Z_q$ and broadcasting $u_i$, $v_i$ and $w_i$. But without knowing the secret shares, $f_S(u_{S_i})$, it is difficult to generate a valid partial signature $s_i$ to satisfy the verification equation,

$$S_S = \sum_{i=1,}^{t} s_i \bmod q \text{ and } g^{S_S} \stackrel{?}{\equiv} R_R \cdot y_S^{R_S} \bmod p.$$

(g). Can one forge a signature $\{S_S, U_S, W_S, m\}$ by the following equation,

$$g^{S_S} \equiv R_R \cdot y_S^{R_S} \bmod p?$$

A forger may randomly selects an integer $R_R$ and then computes the hash value $R_S$ such that $R_S = h(R_R, m) \bmod q$.

Obviously, to computes the integer $S_S$ is equivalent to solving the discrete logarithm problem. On the other hand, the forger can randomly select $R_R$ and $S_S$ first and then try to determine a value $R_S$, that satisfy the equation

$$g^{S_S} \equiv R_R \cdot y_S^{R_S} \bmod p.$$

However, according to the one-way property of the hash function $h$, it is quite impossible. Thus, this attack will not be successful.

(h). Can $t$ or more shareholders act in collusion to reconstruct the polynomial $f_S(x)$ ?

According to the equation $f_S(x) = \sum_{i=1}^{t} f_S(u_{S_i}) \prod_{j=1, j \neq i}^{t} \frac{x - u_{S_j}}{u_{S_i} - u_{S_j}} \bmod q$, the secret polynomial $f_S(x)$ can be reconstructed with the knowledge of any $t$ secret shares, $f_S(u_{S_i})$, $i \in G_S$. So if in an organization the shareholders are known to each other, any $t$ of them may collude and find the secret polynomial $f_S$. This attack, however, does not weaken the security of our scheme in the sense that the number of users that have to collude in order to forge the signature is not smaller than the threshold value.

## Illustration

For illustration, Suppose $|G_S|=7$, $|H_S|=4$, $|G_R|=6$ and $|H_S|=5$. We take the parameters $p = 47$, $q = 23$ and $g = 2$.

**Group Secret Key and Secret Shares Generation for the Organization S**

(a). CTC selects for the group $G_S$, a polynomial $f_S(x) = 11 + 3x + 13x^2 + x^3 \mod 23$.

$$\text{Here } x_S = 11 \text{ and } y_S = 2^{11} \mod 47 = 27.$$

(b). CTC randomly selects $K = 8$ and computes a public value $W = 9$.

(c). CTC computes the secret key, public key, secret share and public value for each member of the group $G_S$, as shown by the following table.

| VALUE \ USER | $y_{S_i}$ | $x_{S_i}$ | $u_{S_i}$ | $f_S(u_{S_i})$ | $v_{S_i}$ |
|---|---|---|---|---|---|
| User – S$_1$ | 7 | 12 | 2 | 8 | 34 |
| User – S$_2$ | 37 | 10 | 9 | 3 | 34 |
| User – S$_3$ | 28 | 14 | 8 | 22 | 38 |
| User – S$_4$ | 18 | 16 | 11 | 4 | 9 |
| User – S$_5$ | 8 | 21 | 3 | 3 | 6 |
| User – S$_6$ | 3 | 19 | 5 | 16 | 6 |
| User – S$_7$ | 36 | 17 | 4 | 19 | 40 |

**Group Secret Key and Secret Shares Generation for the Organization R**

(a). CTC selects for the group $G_R$ a polynomial $f_R(x) = 7 + 2x + 4x^2 + 3x^3 \mod 23$.

$$\text{Here } x_R = 7 \text{ and } y_R = 2^7 \mod 47 = 34.$$

(b). CTC computes the secret key, public key, secret share and public value for each member of the group $G_R$, as shown by the following table.

| VALUE USER | $y_{R_i}$ | $x_{R_i}$ | $u_{R_i}$ | $f_R(u_{R_i})$ | $v_{R_i}$ |
|---|---|---|---|---|---|
| User – R₁ | 9 | 15 | 11 | 21 | 14 |
| User – R₂ | 42 | 9 | 5 | 9 | 25 |
| User – R₃ | 27 | 11 | 8 | 21 | 16 |
| User – R₄ | 25 | 18 | 3 | 15 | 20 |
| User – R₅ | 17 | 6 | 6 | 6 | 24 |
| User – R₆ | 14 | 13 | 4 | 18 | 32 |

## Signature generation by any *t* users

If S₂, S₄, S₆ and S₇ are agree to sign a message *m* for the organization *R*, then

(a). S₂ selects $K_{2_1} = 5$, $K_{2_2} = 7$ and computes $u_2 = 18$, $v_2 = 32$ and $w_2 = 21$.

(b). S₄ selects $K_{4_1} = 4$, $K_{4_2} = 3$ and computes $u_4 = 6$, $v_4 = 16$ and $w_4 = 4$.

(c). S₆ selects $K_{6_1} = 12$, $K_{6_2} = 18$ and computes $u_6 = 32$, $v_6 = 7$ and $w_6 = 1$.

(d). S₇ selects $K_{7_1} = 21$, $K_{7_2} = 11$ and computes $u_7 = 7$, $v_7 = 12$ and $w_7 = 17$.

(e). Each users computes $U_S = 34$, $V_S = 3$, $W_S = 18$ and $R_S = h(3, m) = 8$ (let).

(f). S₂ recovers $f_S(u_{S_2}) = 3$ and computes $MS_{S_2} = 5$ and $s_2 = 22$.

(g). S₄ recovers $f_S(u_{S_4}) = 4$ and computes $MS_{S_4} = 21$ and $s_4 = 11$.

(i). S₆ recovers $f_S(u_{S_6}) = 16$ and computes $MS_{S_6} = 12$ and $s_6 = 16$.

(j). S₇ recovers $f_S(u_{S_7}) = 19$ and computes $MS_{S_7} = 19$ and $s_7 = 12$.

(k). CTC produces a group signature $S_S = 15$ and sends $\{15, 34, 18, m\}$ to the designated combiner *DC* of organization *R* as signature of the group *S* for the message *m*.

## Signature Verification by the Organization *R*

Suppose five users R₁, R₃, R₄, R₅ and R₆ want to verify the signature, then

(a). R1 recovers $f_R(u_{R_1}) = 21$ and computes $MS_{R_1} = 19$.

(b). $R_3$ recovers $f_R(u_{R_3}) = 21$ and computes $MS_{R_3} = 4$.

(c). $R_4$ recovers $f_R(u_{R_4}) = 15$ and computes $MS_{R_4} = 11$.

(d). $R_5$ recovers $f_R(u_{R_5}) = 12$ and computes $MS_{R_5} = 9$.

(e). $R_6$ recovers $f_R(u_{R_6}) = 18$ and computes $MS_{R_6} = 10$.

(f). $DC$ computes the value $R_R = 3$ and then recovers $R_S = h(3, m) = 8$.

(g). $DC$ checks the congruence $2^{15} \stackrel{?}{\equiv} 3 \cdot 27^8 \mod 45$ for a valid signature. This congruence holds so $\{15, 34, 18, m\}$ is a valid signature on the message $m$.

## 4.4. Remarks

In this chapter, we have proposed a **threshold signature scheme with threshold verification.** To obtain this construction, we have used the ElGamal public key cryptosystem and Schnorr's signature scheme. The security of this cryptosystem is based on the discrete log problem. The signature generation is done by certain designated sub – groups of signers and the verification is done by certain designated sub-groups of the group of the receivers. Here designated sub – groups are characterized by threshold values. The threshold value can be different for signature generation and for signature verification. Until $t$ (the threshold value of the group of senders) individuals act in collusion they will get no information about the group secret key. Similarly, until $k$ (the threshold value of the group of recipients) individuals act in collusion they will get no information about the group secret key. The group public parameter $p, q, g$ and a collision free one-way hash function $h$ is same for both the organizations.

In any case, of dispute between the group $S$ and $R$, the CTC keeps the records of signatures and plays the role of a trusted judge. Since the CTC can checks the validity of the signature so when any third party needs the signature verification, the CTC convince the third party about the facts.

# Chapter 5
# Directed-Threshold Multi-Signature Scheme

# Directed-Threshold Multi-Signature Scheme

## 5.1. Introduction

*In threshold schemes*, when any $t$ or more shareholders act in collusion, they can impersonate any other set of shareholders to forge the signatures. In this case, the malicious set of signers does not have any responsibility for the signatures and it is also impossible to trace the signers. Unfortunately, with the ($t$, $n$) threshold schemes proposed so far, this problem can not be solved.

*In multisignature schemes,* the signers of a multisignature are identified in the beginning and the validity of the multisignature has to be verified with the help of identities of the signers. For multisignatures, it is indeed unnecessary to put a threshold value to restrict the number of signers. Consider the situation, where a group of anonymous members would have to generate a multisignature. The members of this group use pseudonyms as their identities in the public directory. What concerns the verifier most is that a message is signed by at least $t$ members and they indeed come from that group. But the verifier has no way to verify whether a user is in fact a member of that group because of the anonymity of the membership. In this case, the multisignature schemes cannot solve this problems, however, the threshold signature schemes do.

On the other hands, there are so many situations, when the signed message is sensitive to the signature receiver and required the approval or consent of several people. Combining all these ideas, we propose a digital signature scheme named as Directed - Threshold Multi - Signature Scheme.

## 5.2. Directed - Threshold Multi - Signature Scheme

In this chapter, we are going to combine the idea of **($t$, $n$) threshold signature schemes and multisignature schemes with directed signature scheme** and propose a new type of signature scheme, called the **Directed - Threshold Multi-Signature Scheme.**

For our construction, we assume that there is a **trusted share distribution center (SDC),** which determine the group secret keys, secret shares of all shareholders, and a **designated combiner** *DC* who takes the responsibility to collect and verify each partial signature and then produce a group signature.

This scheme consists of the following steps.

## 5.2.1. Group Secret Key and Secret Shares Generation for the Organization *S*

(a). **SDC** selects the group public parameters *p, q, g* and a collision free one way hash function *h*. SDC also selects a polynomial

$$f_S(x) = a_0 + a_1 x + \ldots a_{t-1} x^{t-1} \mod q, \text{ with } a_0 = x_S = f_S(0).$$

(b). **SDC** computes the group public key, $y_S = g^{f_S(0)} \mod p$.

(c). **SDC** randomly selects $K \in Z_q$ and computes a public value $W = g^{-K} \mod p$.

(d). **SDC** randomly selects $K_i \in Z_q$ and computes $l_i = [K_i + f_S(u_{S_i})] \mod q$.

   Here $u_{S_i}$ is the public value associated with each user *i* in the group *Gs*.

(e). SDC computes a public value $v_{S_i}$ for each member of the group $G_S$, as,

$$v_{S_i} = l_i \cdot y_{S_i}^K \mod p.$$

   Here $y_{S_i}$ is the public value associated with each user *i* in the group *Gs*.

(f). SDC sends $\{v_{S_i}, W\}$ to each user *i* in the group $G_S$ through a public channel.

(g). SDC also needs to computes public values, $m_i$ and $n_i$, as,

$$m_i = g^{l_i} \mod p \text{ and } n_i = g^{K_i} \mod p.$$

## 5.2.2. Partial Signature Generation By Any $t$ Users and Verification

If any $t$ members of the organization $S$ out of $n$ members agree to sign a message $m$ for a person $R$. $R$ possesses a pair $(x_R, y_R)$. Then the signature generation has the following steps.

(a) Each user $i \in H_S$ randomly selects $K_{i_1}$ and $K_{i_2} \in Z_q$ and computes

$$u_i = g^{-K_{i_2}} \bmod p, \quad v_i = g^{K_{i_1}} \bmod p \text{ and } w_i = g^{K_{i_1}} y_R^{K_{i_2}} \bmod p.$$

(b) Each user makes $u_i$, $w_i$ publicly and $v_i$ secretly available to each member $i \in H_S$
Once all $u_i$, $v_i$ and $w_i$ are available, each member $i \in H_S$ computes the product $U_S$, $V_S$, $W_S$ and a hash value $R_S$, as,

$$U_S = \prod_{i \in H_S} u_i \bmod q, \quad V_S = \prod_{i \in H_S} v_i \bmod q,$$

$$W_S = \prod_{i \in H_S} w_i \bmod q \text{ and } R_S = h(V_S, m) \bmod q.$$

(c). Each user $i \in H_S$ recovers his/her secret share $l_i = v_{S_i} W^{x_{S_i}} \bmod p$.

(d). Each user $i \in H_S$ modifies his/her shadow $MS_{S_i} = l_i \cdot \prod_{j=1, j \neq i}^{t} \frac{-u_{S_j}}{u_{S_i} - u_{S_j}} \bmod q$.

(e). Each user $i \in H_S$ uses his/her modified shadow $MS_{S_i}$ and computes a value $s_i$, as,

$$s_i = K_{i_1} + MS_{S_i} \cdot R_S \bmod q.$$

(e) Each member $i \in H_S$ sends his partial signature to the $DC$.

(f) The $DC$ checks the congruence $g^{s_i} \stackrel{?}{\equiv} v_i \cdot m_i \left( \prod_{j \in H_S, j \neq i} \frac{-u_{S_i}}{u_{S_i} - u_{S_j}} \bmod q \right)^{R_S} \bmod p$ for a partial valid signature ($s_i$, $v_i$, $R_S$). If this congruence holds, then the partial signature ($s_i$, $v_i$, $R_S$) for shareholder $i$ is valid.

## 5.2.3. Group Signature Generation

(a). **DC** can computes the group signature $S_S = \sum_{i=1,}^{t} s_i \bmod q$ by combining all the partial signature.

(b). **DC** sends $\{ S_S, U_S, W_S, m\}$ to $R$ as signature of the group $S$ on the message $m$.

## 5.2.4. Signature Verification by R

To verify the validity of the group signature $\{S_S, U_S, W_S, m\}$ **the verifier $R$ needs his/her secret key** $x_R$.

(a) The verifier $R$ computes a verification value $E$, as,

$$E = \prod_{i \in H_S} n_i \left( \prod_{j \in H_S, j \neq i} \frac{0 - u_{S_j}}{u_{S_i} - u_{S_j}} \bmod q \right) \bmod p.$$

(b). The verifier $R$ can recovers the values $R_R$ and $R_S$, as

$$R_R = W_S . U_S^{x_R} \bmod p \quad \text{and} \quad R_S = h(R_R, m).$$

(c). The verifier $R$ uses the concurrence $g^{S_S} \stackrel{?}{\equiv} R_R . (E . y_S)^{R_S} \bmod p$ to check the validity of the signature. If this concurrence holds, then the group signature $\{S_S, U_S, W_S, m\}$ is valid signature of the organization $S$ on the message $m$.

## 5.2.5. Proof of Validity By R to any third Party C

(a). $R$ computes $\mu = U_S^{x_R} \bmod p$ and $R_R = \mu . W_S \bmod p$.

(b) $R$ sends $\{ R_R, E, S_S, U_S, m, \mu\}$ to C.

(c) C recovers $R_S = h(R_R, m)$ and uses the following concurrence to check the validity of the signature

$$g^{S_S} \stackrel{?}{\equiv} R_R . (E . y_S)^{R_S} \bmod p.$$

If this does not hold C stops the process; otherwise goes to the next steps.

(d) $R$ in a zero knowledge fashion proves to C that $\log_{U_S} \mu = \log_g y_R$ as follows.

This part of the scheme runs as in section - 2.2.4

## 5.3. Security Discussions

In this sub-section, we shall discuss the security aspects of proposed ***Directed - threshold– multisignature scheme***. Here we shall discuss several possible attacks. But none of these can successfully break our system.

**(a).** Can any one retrieve the secret keys $x_S = f_S(0)$ from the group public key $y_S$ ?

This is as difficult as solving discrete logarithm problem. No one can get the secret key $x_S$, since $f_S$ is the randomly and secretly selected polynomials by the SDC. On the other hand, by using the public keys $y_S$ no one also get the secret keys $x_S$ because this is as difficult as solving discrete logarithm problem.

**(b).** Can one retrieve the secret shares, $f_S(u_{S_i})$ $i \in G_S$, from the equation

$$v_{S_i} = f_S(u_{S_i}) \cdot y_{S_i}^K \mod p \ ?$$

No because $f_S$ is a randomly and secretly selected polynomial and $K$ is also a randomly and secretly selected integer by the SDC.

**(c).** Can one retrieve the secret shares, $f_S(u_{S_i})$ $i \in G_S$, from the equation

$$f_S(u_{S_i}) = v_{S_i} W^{x_{S_i}} \mod p \ ?$$

Only the user $i$ can recovers his secret shares, $f_S(u_{S_i})$, because $f_S$ is a randomly and secretly selected polynomial and $x_{S_i}$ is secret key of the user $i \in G_S$.

**(d).** Can one retrieve the modified shadow $MS_{S_i}$, integer $K_{i_1}$, $R_S$ and partial signature $s_i, i \in G_S$ from the equation $s_i = K_{i_1} + MS_{S_i} \cdot R_S \mod q$ ?

It is computationally infeasible for a forger to collect the $MS_{S_i}$, integer $K_{i_1}$, $R_S$ and partial signature $s_i, i \in G_S$.

**(e).** Can the designated combiner $DC$ retrieve the any partial information from the equation, $S = \sum_{i=1,}^{t} s_i \bmod q$ ?

Obviously, this is computationally infeasible for $DC$.

**(f).** Can one impersonate a user $i \in H$ ?

A forger may try to impersonate a user $i \in H_S$, by randomly selecting two integers $K_{i_1}$ and $K_{i_2} \in Z_q$ and broadcasting $u_i$, $v_i$ and $w_i$. But without knowing the secret shares, $f_S(u_{S_i})$ and $R_S$, it is difficult to generate a valid partial signature $s_i$ to satisfy the verification equations,

$$g^{s_i} \stackrel{?}{\equiv} v_i \cdot m_i \left( \prod_{j \in H_S, j \neq i} \frac{-u_{S_i}}{u_{S_i} - u_{S_j}} \bmod q \right)^{R_S} \bmod p.$$

**(g).** Can one forge a signature $\{S_S, U_S, W_S, m\}$ by the following equation,

$$g^{S_S} \stackrel{?}{\equiv} R_R \cdot (E \cdot y_S)^{R_S} \bmod p \ ?$$

A forger may randomly selects an integer $R_R$ and then computes the hash value $R_S$ such that $R_S = h(R_R, m) \bmod q$.

Obviously, to computes the integer $S_S$ is equivalent to solving the discrete logarithm problem. On the other hand, the forger can randomly select $R_S$ and $S_S$ first, then try to determine a value $R_R$ that satisfy the equation

$$g^{S_S} \stackrel{?}{\equiv} R_R \cdot (E \cdot y_S)^{R_S} \bmod p.$$

However, according to the property of the hash function $h$, it is quite impossible. Thus, this attack will not be successful.

**(h).** Can $t$ or more shareholders act in collusion to reconstruct the polynomial $f_S(x)$ ?

According to the equation,

$$f_S(x) = \sum_{i=1}^{t} f(u_{S_i}) \prod_{j=1, j \neq i}^{t} \frac{x - u_{S_j}}{u_{S_i} - u_{S_j}} \bmod q,$$

the secret polynomial $f_S$ can be reconstructed with the knowledge of any $t$ secret shares, $f_S(u_{S_i})$ $i \in G_S$.

But in our proposed scheme, the secret shares $l_i$, $i \in G_S$, contains the integer $K_i$ which known only by the trusted SDC and has to be removed first before reconstructing the polynomial $f_S(x)$. A malicious shareholder $i$ may try to retrieve the integer $K_i$ from the public key $n_i$. However, the difficulty is same as solving the discrete logarithm problem. Thus, any $t$ or more shareholders cannot conspire to reconstruct the polynomial $f_S(x)$ by providing their own secret shares.

So if in an organization the shareholders are known to each other, even than they cannot reconstruct the polynomial $f_S(x)$.

## Illustration

For illustration, suppose $|G_S| = 7$, $|H_S| = 5$, $p = 47$, $q = 23$, $g = 25$.

| VALUE<br>USER | Secret<br>$x_i$ | Public<br>$y_i$ | Secret<br>$K_i$ | Secret<br>$f_S(u_{S_i})$ | Secret<br>$l_i$ | Public<br>$m_i$ | Public<br>$n_i$ | Public<br>$v_i$ |
|---|---|---|---|---|---|---|---|---|
| User–S$_1$ | 13 | 16 | 8 | 2 | 10 | 3 | 17 | 41 |
| User–S$_2$ | 18 | 4 | 22 | 21 | 20 | 9 | 32 | 29 |
| User–S$_3$ | 19 | 6 | 10 | 16 | 3 | 21 | 3 | 1 |
| User–S$_4$ | 20 | 9 | 17 | 3 | 20 | 9 | 34 | 19 |
| User–S$_5$ | 17 | 34 | 14 | 14 | 5 | 12 | 24 | 38 |
| User–S$_6$ | 22 | 32 | 16 | 14 | 7 | 27 | 7 | 14 |
| User–S$_7$ | 15 | 36 | 21 | 22 | 20 | 9 | 37 | 44 |

**Group Secret Key and Secret Shares Generation for the Organization S**

(a). SDC selects a polynomial $f_S(x) = 13 + 18 x^4$ mod 23, so $x_S = 13$ and $y_S = 16$.

(b). SDC randomly selects $K = 14$ and computes a public value $W = 2$.

(c). SDC computes the secret key, public key, secret share and public value for each member of the group $G_S$, as shown in the above table.

## Signature Generation by any $t$ Users

If any five users $S_2$, $S_4$, $S_5$, $S_6$, and $S_7$ out of seven members agree to sign a message $m$, for a person $R$ possesses $x_R = 9$, $y_R = 2$ then the signature generation has the following steps.

(a). $S_2$ randomly selects $K_{2_1} = 18$, $K_{2_2} = 17$ and computes $u_2 = 18$, $v_2 = 4$ and $w_2 = 3$.

(b). $S_4$ randomly selects $K_{4_1} = 17$, $K_{4_2} = 19$ and computes $u_4 = 8$, $v_4 = 34$ and $w_4 = 8$.

(c). $S_5$ randomly selects $K_{5_1} = 14$, $K_{5_2} = 13$ and computes $u_5 = 3$, $v_5 = 24$ and $w_4 = 7$.

(d). $S_6$ randomly selects $K_{6_1} = 19$, $K_{6_2} = 21$ and computes $u_6 = 14$, $v_6 = 6$ and $w_6 = 25$.

(e). $S_7$ randomly selects $K_{7_1} = 16$, $K_{7_2} = 18$ and computes $u_7 = 12$, $v_7 = 7$ and $w_7 = 34$.

(f). Each user computes the product $U_S = 8$, $V_S = 36$, $W_S = 14$ and $R_S = 9$ (let).

(g). $S_2$ recovers his/her secret share $l_2 = 20$ and computes $MS_{S_2} = 10$, $s_2 = 16$.

(h). $S_4$ recovers his/her secret share $l_4 = 20$ and computes $MS_{S_4} = 16$, $s_4 = 0$.

(i). $S_5$ recovers his/her secret share $l_5 = 5$ and computes $MS_{S_5} = 6$, $s_5 = 22$.

(j). $S_6$ recovers his/her secret share $l_6 = 7$ and computes $MS_{S_6} = 5$, $s_6 = 18$.

(k). $S_7$ recovers his/her secret share $l_7 = 20$ and computes $MS_{S_7} = 5$, $s_7 = 15$.

## Partial Signature Verification and Signature Generation by DC

(a). $DC$ verifies each partial signature. For example for user $S_2$,

$$s_2 = 16, \; v_2 = 4, \; m_2 = 9, \; R_S = 9, \; \prod_{j=4,5,6,7. i=2} \frac{-u_{S_j}}{u_{S_i} - u_{S_j}} \mod 23 = 12.$$

He checks $25^{16} \stackrel{?}{\equiv} 4 \cdot \left(9^{12}\right)^9 \mod 47$. This holds. Similarly, he checks each partial signature.

(b). *DC* computes a group value $S_S = 2$ and sends $\{2, 14, 8, m\}$ as signature of the group *S* for the message *m*.

### Signature Verification by the User *R*.

(a). The verifier *R* computes a verification value $E = 18$.

(b). The verifier *R* can recovers the values $R_R = 36$ and $R_S = 9$.

(c). The verifier *R* uses the congruence $25^2 \stackrel{?}{\equiv} 36 \cdot (18.16)^9 \mod 47$ to check the validity of the signature

This congruence holds, so the group signature $\{2, 14, 8, m\}$ is valid signature of the group *S* on the message *m* for the person *R*.

### Proof of Validity by *R* To any third Party C.

(a). *R* computes $\mu = 16$, $R_R = 36$ and sends $\{36, 18, 2, 8, m, 16\}$ to C.

(b). C recovers $R_S = 9$ and check the concurrence $25^2 \stackrel{?}{\equiv} 36 \cdot (18.16)^9$ for the validity of the signature. This holds; so C goes to the next steps.

(c). *R* in a zero knowledge fashion proves to C that $\log_8 16 = \log_{25} 2$ as follows:-

- C chooses random $u = 9$, $v = 11$ and computes $w = 25$ and sends *w* to *R*.
- *R* chooses random $\alpha = 37$ computes $\beta = 36$, $\gamma = 9$ and sends $\beta$, $\gamma$ to C.
- C sends *u, v* to *R*, by which *R* can verify that $w = 25$.
- *R* sends $\alpha$ to C, by which she can verify that $\beta = 36$ and $\gamma = 9$.

## 5.4. Remarks

In this chapter, we have proposed a **Directed Threshold – Multisignature Scheme**. In this scheme,

- There is a **trusted SDC** that determine the group secret keys, all shareholder's secret shares
- There is a **designated combiner DC** who takes the responsibility to collect and verify each partial signature and then produce a group signature, **but no secret information associated with the DC**

- Any malicious set of signers cannot impersonate any other set of signers to forge the signatures. In case of forgery, it is possible to trace the signing set.

- Any $t$ or more shareholders act in collusion cannot conspire to reconstruct the polynomial $f_S(x)$ by providing their own secret shares and hence they cannot recover the group secret key.

# Chapter 6

# Directed-Threshold Multi-Signature Scheme without SDC

# Directed-Threshold Multi-Signature Scheme without SDC

## 6.1. Introduction

Many threshold signature schemes require a **trusted SDC** to generate the group secret keys and secret shares of group members, which have a single point of vulnerability. The existence of such a center is not a reasonable assumption; there are two potential problems.

- First, for many applications, there is no one person or devices which cab be completely trusted by all members of the group.
- Second, the use of a key center creates a single point failure. Any security lapse at the key center can reveal the private key.

To avoid these problems, in 1992, Harn introduced a scheme based on a modified ELGamal signature scheme, which does not require a trusted SDC [42,61]. Each member works as a SDC, generates and distributes the secret shares for each user. In this chapter, we proposed a *Directed - Threshold Multi - Signature Scheme without SDC.*

## 6.2. Directed - Threshold Multi - Signature Scheme without SDC

In this scheme, each shareholder works as a SDC to generate his secret key and distribute the corresponding secret shares to other shareholders. In advance, all the shareholders are agree for the public parameters. We assume that there is a **designated combiner** *DC* who takes the responsibility to collect and verify each partial signature and then produce a group signature. **Nevertheless, there is no partial secret information of the other users associated with the *DC*.**

### 6.2.1. Group Public Key and Secret Shares Generation Phase

(a). All the members of the organization $S$ agree to select the group public parameters $p, q, g, h$ and a common random secret $K$.

(b). Each member $i$ in $G_S$ computes $W = g^{-K} \bmod p$.

(c). Each member $i$ in $G_S$ randomly selects a $(t-1)^{th}$ degree polynomial $f_i(x)$ secretly and an integer $u_{S_i} \in Z_q$ publicly.

(d). Each member $i$ in $G_S$ computes partial group public key $y_i = g^{f_i(0)} \mod p$.

(e). The group public key $y_S$ is given by $y_S = \prod_{i \in G_S} y_i \mod p$.

(f). Each member $i$ in $G_S$ works as SDC. He selects a random secret $h_{ij} \in Z_q$ and computes the secret value $l_{ij}$ and public value $m_{ij}$, $n_{ij}$ for $j \neq i$, as,

$$l_{ij} = [h_{ij} + f_i(u_{S_i})] \mod q,$$

$$m_{ij} = g^{l_{ij}} \mod p \text{ and } n_{ij} = g^{h_{ij}} \mod p.$$

(g). Each member $i$ in $G_S$ computes a public value $v_{S_i}$ for each member $j \neq i$, of the group $G_S$, as, $v_{S_{ij}} = l_{ij} \cdot y_{S_j}^K \mod p$.

Here, $y_{S_j}$ is the public key associated with each user $j$ in the group $G_S$.

(h). Each member $i$ in $G_S$ sends $v_{S_{ij}}$ to each member $j \neq i$, through a public channel.

## 6.2.2. Partial Signature Generation by any $t$ Members and Verification

If any $t$ members of the organization out of $n$ members agree to sign a message $m$ for a person R. R possesses a pair $(x_R, y_R)$. Then the signature generation has the following steps.

(a). Each member $i, i \in H_S$, randomly selects $K_{i_1}$ and $K_{i_2} \in Z_q$ and computes

$$u_i = g^{-K_{i_2}} \mod p, \quad v_i = g^{K_{i_1}} \mod p \text{ and } w_i = g^{K_{i_1}} y_R^{K_{i_2}} \mod p.$$

(b). Each member makes $u_i$, $w_i$ publicly and $v_i$ secretly available to each member of $H_S$. Once all $u_i, v_i$ and $w_i$ are available, each member $i, i \in H_S$ computes the product $U_S$, $V_S$, $W_S$ and a hash value $R_S$, as,

$$U_S = \prod_{i \in H_S} u_i \mod q, \quad V_S = \prod_{i \in H_S} v_i \mod q,$$

$$W_S = \prod_{i \in H_S} w_i \bmod q, \text{ and } R_S = h(V_S, m) \bmod q.$$

(c). Each member $i, i \in H_S$ recovers his/her secret share $l_{ji}$, as,

$$l_{ji} = v_{S_{ji}} W^{x_{S_i}} \bmod p. \quad (j \neq i.)$$

(d). Each member $i, i \in H_S$ computes a value $C_i = \prod_{k \in H_S, k \neq i} \frac{(0 - u_{S_k})}{(u_{S_i} - u_{S_k})} \bmod q.$

(e). Each member of $H_S$ computes his/her modified shadow $MS_{S_i} = \sum_{j \in G_S, j \notin H_S} l_{ji} \cdot C_i \bmod q.$

(f). Each member $i, i \in H_S$ uses his/her modified shadow, $MS_{S_i}$ and a value $s_i$, as

$$s_i = [K_{i_1} + (f_i(0) + MS_{S_i}) \cdot R_S] \bmod q.$$

(g). Each member $i, i \in H_S$ sends his partial signature to the designated combiner $DC$. $DC$ verify the partial signature ($s_i, v_i, C_i, R_S$) by the relation,

$$g^{s_i} \stackrel{?}{\equiv} v_i \cdot \left( y_i \prod_{j \in G_S, j \notin H_S} m_{ji}^{C_i} \right)^{R_S} \bmod p.$$

If the above equation holds, then the partial signature ($s_i, v_i, C_i, R$) for shareholder $i$ is valid.

## 6.2.3. Group Signature Generation.

(a). **DC** can computes the group signature $S_S = \sum_{i=1}^{t} s_i \bmod q$ by combining all the partial signature.

(b). **DC** sends $\{S_S, U_S, W_S, m\}$ as signature of the group $S$ for the message $m$ to $R$.

## 6.2.4. Signature Verification by *R*

To verify the validity of the group signature $\{S_S, U_S, W_S, m\}$ **the verifier R needs is/her secret key** $x_R$. This sub-section consists of the following steps.

(a). The verifier $R$ computes a verification value $E = \prod_{i \in H_S} \left( \left( \prod_{j \in G_S, j \notin H_S} n_{ji} \right)^{C_i} \right) \bmod p.$

(b). Only the verifier $R$ can recovers the values $R_R$ and $R_S$, as

$$R_R = W_S \cdot U_S^{x_R} \bmod p \text{ and } R_S = h(V_S, m).$$

(c). The verifier $R$ uses the congruence $g^{S_S} \stackrel{?}{\equiv} R_R \cdot (E \cdot y_S)^{R_S} \bmod p$ to check the validity the signature. If this congruence holds, then the group signature $\{S_S, U_S, W_S, m\}$ is valid signature of the organization $S$ on the message $m$.

## 6.2.5. Proof of Validity by $R$ to any third Party C

This part of the protocol is same as in section 5.2.5

## 6.3. Security Discussions

In this sub-section, we shall discuss the security aspects of proposed scheme. Here we shall discuss several possible attacks and show that, none of these can successfully break the system.

(a). Is it possible to retrieve the partial secret keys $f_i(0)$, $i \in G_S$ ?

This is as difficult as solving discrete logarithm problem. No one can get the partial group public key $y_i$, since $f_i$ is the randomly and secretly selected polynomial by the member $i$. On the other hand, by using the public keys $y_S$ no one also get the partial secret keys $f_i(0)$ because

$$y_i = g^{f_i(0)} \bmod p \text{ and } y_S = \prod_{i \in G_S} y_i \bmod p.$$

(b). Is it possible to retrieve the secret share $l_{ij}$ from the equation

$$l_{ij} = [h_{ij} + f_i(u_{S_i})] \bmod q ?$$

No, because $f_i$ is the randomly and secretly selected polynomial and $h_{ij}$ is also a randomly and secretly selected integer by the member $i$.

(c). Is it possible to retrieve the secret share $l_{ij}$ from the equation

$$m_{ij} = g^{l_{ij}} \bmod p ?$$

No, because this is as difficult as solving discrete logarithm problem.

(d). Is it possible to retrieve the secret share $l_{ij}$ from the equation

$$v_{S_{ij}} = l_{ij} \cdot y_{S_j}^{K} \bmod p?$$

No because $K$ is a randomly and secretly selected common integer.

(e). Is it possible to retrieve the secret shares $l_{ij}$ from the equation

$$l_{ji} = v_{S_{ji}} W^{x_{S_i}} \bmod p?$$

Only the user $i$ can recovers his secret shares $l_{ij}$ because $x_{S_i}$ is secret key of the user $i$.

(f). Can one retrieve the modified shadow $MS_{S_i}$ from the equation

$$MS_{S_i} = \sum_{j \in G_S, j \notin H_S} l_{ji} \cdot C_i \bmod q \ ?$$

It is impossible to collect the modified shadow $MS_{S_i}$ from the equation because all $l_{ij}$ are secret information shared by the users.

(g). Is it possible that the designated combiner $DC$ retrieve the any partial information from the equation $S_S = \sum_{i \in H_S} s_i \bmod q \ ?$

Obviously, this is computationally infeasible for $DC$.

(h). Is it possible to that any one can impersonate a user $i \in H$ ?

A forger may try to impersonate a user $i \in H_S$, by randomly selecting two integers $K_{i_1}$ and $K_{i_2} \in Z_q$ and broadcasting $u_i$, $v_i$ and $w_i$. But without knowing the secret shares $l_{ij}$ and $R_S$, it is difficult to generate a valid partial signature $s_i$ to satisfy the verification equations, $g^{s_i} \overset{?}{\equiv} v_i \cdot \left( y_i \prod_{j \in G_S, J \notin H_S} m_{ji}^{C_i} \right)^{R_S} \bmod p.$

(g). Is it possible to that any one can forge a signature $\{S_S, U_S, W_S, m\}$ by the following equation $g^{S_S} \overset{?}{\equiv} R_R \cdot (E \cdot y_S)^{R_S} \bmod p$?

A forger may randomly selects an integer $R_R$ and then computes the hash value $R_S$ such that $R_S = h(R_R, m) \bmod q$, obviously to compute the integer $S_S$ is equivalent to solving the discrete logarithm problem. On the other hand, the forger can randomly select $R_S$ and $S_S$ first, then try to determine a value $R_R$ that satisfies the signature verification equation. However, according to the property of the hash function $h$, it is quite impossible. Thus, this attack will not be successful.

# Illustration

For illustration, Suppose $|G_S| = 7$, $|H_S| = 4$, $p = 47$, $q = 23$, $g = 3$, $K = 11$ and $W = 12$.

## Group public Key and Secret Shares Generation Phase.

(a). All the users compute their secret and public values, which are given by the following table.

| VALUE USER | Secret $f_i(x)$ | Secret $f_i(0)$ | Public $u$ | Public $y_i$ | Secret $x_{S_i}$ | Public $y_{S_i}$ |
|---|---|---|---|---|---|---|
| User-$S_1$ | $7 + 12\ x^3$ | 7 | 9 | 25 | 9 | 37 |
| User-$S_2$ | $9 + 11\ x^3$ | 9 | 13 | 37 | 11 | 4 |
| User-$S_3$ | $14 + 8\ x^3$ | 14 | 15 | 14 | 13 | 36 |
| User-$S_4$ | $17 + 3\ x^3$ | 17 | 11 | 2 | 19 | 18 |
| User-$S_5$ | $13 + 7\ x^3$ | 13 | 18 | 36 | 5 | 8 |
| User-$S_6$ | $18 + 15x^3$ | 18 | 19 | 6 | 10 | 17 |
| User-$S_7$ | $21 + 15\ x^3$ | 21 | 21 | 21 | 14 | 14 |

(b). The group public key $y_S$ is given by $y_S = 25$.

(c). User-$S_1$ makes the following table.

| VALUE USER | secret $h_{1j}$ | secret $h_{1j}$ | public $m_{1j}$ | public $n_{1j}$ | public $v_{1j}$ |
|---|---|---|---|---|---|
| User – $S_2$ | 14 | 4 | 34 | 14 | 2 |
| User – $S_3$ | 9 | 13 | 36 | 37 | 10 |
| User – $S_4$ | 11 | 5 | 8 | 4 | 45 |
| User – $S_5$ | 15 | 17 | 2 | 42 | 18 |
| User – $S_6$ | 17 | 15 | 42 | 2 | 43 |
| User – $S_7$ | 13 | 16 | 32 | 36 | 42 |

(d). User- $S_2$ makes the following table.

| VALUE USER | Secret $h_{2j}$ | Secret $l_{2j}$ | Public $m_{2j}$ | Public $n_{2j}$ | Public $v_{2j}$ |
|---|---|---|---|---|---|
| User – $S_1$ | 13 | 14 | 14 | 36 | 21 |
| User – $S_3$ | 14 | 3 | 27 | 14 | 24 |
| User – $S_4$ | 9 | 8 | 28 | 17 | 25 |
| User – $S_5$ | 7 | 21 | 21 | 25 | 25 |
| User – $S_6$ | 16 | 11 | 4 | 8 | 19 |
| User – $S_7$ | 3 | 16 | 32 | 27 | 42 |

(e). User- $S_3$ and makes the following table.

| VALUE USER | Secret $h_{3j}$ | Secret $l_{3j}$ | Public $m_{3j}$ | Public $n_{3j}$ | Public $v_{3j}$ |
|---|---|---|---|---|---|
| User – $S_1$ | 7 | 11 | 4 | 25 | 40 |
| User – $S_2$ | 10 | 5 | 8 | 17 | 26 |
| User – $S_4$ | 11 | 1 | 3 | 4 | 9 |
| User – $S_5$ | 13 | 16 | 32 | 36 | 28 |
| User – $S_6$ | 19 | 4 | 34 | 18 | 24 |
| User – $S_7$ | 21 | 17 | 2 | 21 | 27 |

(f). User- $S_4$ makes the following table.

| VALUE USER | Secret $h_{4j}$ | Secret $l_{4j}$ | Public $m_{4j}$ | Public $n_{4j}$ | Public $v_{4j}$ |
|---|---|---|---|---|---|
| User – $S_1$ | 19 | 15 | 14 | 18 | 46 |
| User – $S_2$ | 13 | 20 | 7 | 36 | 10 |
| User – $S_3$ | 11 | 10 | 17 | 4 | 33 |
| User – $S_5$ | 6 | 16 | 32 | 24 | 28 |
| User – $S_6$ | 8 | 17 | 2 | 28 | 8 |
| User – $S_7$ | 18 | 11 | 4 | 6 | 23 |

(g). User- $S_5$ makes the following table.

| VALUE USER | Secret $h_{5j}$ | Secret $l_{5j}$ | Public $m_{5j}$ | Public $n_{5j}$ | Public $v_{5j}$ |
|---|---|---|---|---|---|
| User – $S_1$ | 6 | 16 | 32 | 24 | 24 |
| User – $S_2$ | 17 | 22 | 16 | 2 | 11 |
| User – $S_3$ | 18 | 15 | 42 | 6 | 26 |
| User – $S_4$ | 13 | 5 | 8 | 36 | 45 |
| User – $S_6$ | 19 | 21 | 21 | 18 | 32 |
| User – $S_7$ | 8 | 11 | 4 | 28 | 23 |

(h). User- $S_6$ makes the following table

| VALUE USER | Secret $h_{6j}$ | Secret $l_{6j}$ | Public $m_{6j}$ | Public $n_{6j}$ | Public $v_{6j}$ |
|---|---|---|---|---|---|
| User – $S_1$ | 2 | 7 | 25 | 9 | 34 |
| User – $S_2$ | 5 | 19 | 18 | 8 | 33 |
| User – $S_3$ | 7 | 4 | 34 | 25 | 32 |
| User – $S_4$ | 11 | 7 | 25 | 4 | 16 |
| User – $S_5$ | 13 | 19 | 18 | 36 | 45 |
| User – $S_7$ | 12 | 2 | 9 | 12 | 17 |

(i). User- $S_7$ makes the following table.

| VALUE USER | Secret $h_{7j}$ | Secret $l_{7j}$ | Public $m_{7j}$ | Public $n_{7j}$ | Public $v_{7j}$ |
|---|---|---|---|---|---|
| User – $S_1$ | 11 | 19 | 18 | 4 | 5 |
| User – $S_2$ | 13 | 7 | 25 | 36 | 27 |
| User – $S_3$ | 15 | 15 | 42 | 42 | 26 |
| User – $S_4$ | 17 | 16 | 32 | 2 | 3 |
| User – $S_5$ | 14 | 8 | 28 | 16 | 14 |
| User – $S_6$ | 12 | 16 | 32 | 12 | 2 |

### Partial Signature Generation by any *t* Users

If any four members $S_2, S_4, S_6$, and $S_7$ are agree to sign a message $m$ for a person $R$, possessing $x_R = 7$, $y_R = 25$, then the signature generation has the following steps.

(a). $S_2$ randomly selects $K_{2_1} = 11, K_{2_2} = 13$ and computes $u_2 = 1, v_2 = 4$ and $w_2 = 17$.

(b). $S_4$ randomly selects $K_{4_1} = 10, K_{4_2} = 12$ and computes $u_4 = 4, v_4 = 17$ and $w_4 = 9$.

(c). $S_6$ randomly selects $K_{6_1} = 14, K_{6_2} = 17$ and computes $u_6 = 24, v_6 = 14$ and $w_6 = 6$.

(d). $S_7$ randomly selects $K_{7_1} = 18, K_{7_2} = 5$ and computes $u_7 = 6, v_7 = 6$ and $w_7 = 25$.

(e). Each user computes $U_S = 16, R_R = 25, W_S = 14$ and $R_S = h(25, m) = 7$. (let).

(f). $S_2$ recovers his shares $l_{12} = 4, l_{32} = 5, l_{52} = 22$ and $C_2 = 1, MS_2 = 8, s_2 = 15$.

(g). $S_4$ recovers his shares $l_{14} = 5, l_{34} = 1, l_{54} = 5$ and $C_4 = 11, MS_4 = 6, s_4 = 10$.

(h). $S_6$ recovers his shares $l_{16} = 15, l_{36} = 4, l_{56} = 21$ and $C_6 = 9, MS_6 = 15, s_6 = 15$.

(i). $S_7$ recovers his shares $l_{17} = 16, l_{37} = 17, l_{57} = 11$ and $C_7 = 3, MS_7 = 17, s_7 = 8$.

### Partial Signature Verification and Signature Generation by *DC*

(a) *DC* verifies each partial signature. For example for user $S_2$, $s_2 = 15, v_2 = 4, m_{12} = 34, m_{32} = 8, m_{52} = 16, y_{S_2} = 37, R_S = 7, C_2 = 12$.

*DC* check $3^{16} \stackrel{?}{\equiv} 4 \cdot 37^7 (34 \cdot 8 \cdot 16)^7 \bmod 47$. This holds. Similarly, he checks other partial signatures.

(b). *DC* computes a group value $S_S = 2$ and sends $\{2, 16, 14, m\}$ as signature of the group *S* for the message $m$.

### Signature Verification by the Person *R*

(a). The verifier $R$ computes a verification value $E = 12$.

(b). The verifier $R$ can recovers the values $R_R = 25$ and $R_S = 7$.

(c). The verifier $R$ uses the congruence $3^2 \stackrel{?}{\equiv} 25 \cdot (12 \cdot 25)^7 \bmod 47$ to check the validity of the signature.

This congruence holds, so the group signature $\{2, 16, 14, m\}$ is valid signature of the group $S$ on the message $m$ for the person $R$.

**Proof of Validity by *R* to any third Party C**

(a). $R$ computes $\mu = 32$, $R_R = 25$ and sends $\{25, 12, 2, 16, m, 32\}$ to C.

(b). C recovers $R_S = 7$ and check the concurrence $3^2 \stackrel{?}{\equiv} 25 \cdot (12.25)^7 \mod 47$ for the validity of the signature. This holds; so, C goes to the next steps.

(c). $R$ in a zero knowledge fashion proves to C that $\log_{16} 32 = \log_3 25$ as follows:-

- C chooses random $u = 13$, $v = 15$ and computes $w = 9$ and sends $w$ to $R$.
- $R$ chooses random $α = 11$ computes $β = 36$, $γ = 16$ and sends $β$, $γ$ to C.
- C sends $u, v$ to $R$, by which $R$ can verify that $w = 9$.
- $R$ sends $α$ to C, by which she can verify that $β = 36$ and $γ = 16$.

## 6.4. Remarks

In this chapter, we have proposed a **Directed –Threshold Multi - signature Scheme without SDC.** In this scheme,

- **Each shareholder works as a SDC** to generate his secret key and distribute the corresponding secret shares to other shareholders.
- There is a **designated combiner *DC*** who takes the responsibility to collect and verify each partial signature and then produce a group signature, **but no secret information is associated with the *DC*.**
- Any malicious set of signers cannot impersonate any other set of signers to forge the signatures. In case of forgery, it is possible to trace the signing set.
- Any $t$ or more shareholders acting in collusion cannot conspire to reconstruct the group secret key by providing their own secret shares.

# Chapter 7

# The Generalized Directed– Threshold Multi – Signature Scheme

# The Generalized Directed-Threshold Multi-Signature Scheme

## 7.1. Introduction

When the message is sensitive to the signature receiver and *requires the approval of some specified subsets of signers*. In this note, we combine the idea of directed signature scheme with threshold multi-signature scheme and propose a threshold digital signature scheme under the title *Generalized Directed - Threshold Multi - Signature Scheme.* The features in this generalized directed- threshold multi - signature scheme are similar to those of the (t ,n) directed threshold- multi – signature scheme as pointed out in the last chapter except that the group signature can be generated only by some specified subsets of members according to the signature policy.

## 7.2. Generalized Directed - Threshold Multi - Signature Scheme

If the message $m$ (sensitive to the signature receiver $R$) is transmitted by an organization $S$ to a person $R$ and the message requires the approval of some specific subsets of signers among the set of signers, then the responsibility of signing the messages needs to be shared by some specified subsets of the signers according to the signing policy. For this situation, we propose a **Generalized Directed - threshold multi - signature scheme.**

Let $G_s$ is the set of $n$ signers belonging to the organization $S$ and $H_s$ denote the specified subset of $t$ signers. The valid group signature can be generated by the cooperation of the signers belongs to $H_s$. We also define that $P(G_s)$ is the set of all specified subsets.

For our construction, we assume that there is a **trusted share distribution center (SDC) that** determines the group secret keys, all shareholder's secret shares, and a **designated combiner** $DC$ who takes the responsibility to collect and verify each partial signature and then produce a group signature. This scheme consists of the following steps.

## 7.2.1. Group Public Key and Secret Shares Generation Phase

(a). **SDC** selects the group public parameters $p, q, g$ and a collision free one way hash function $h$. SDC also selects a group secret key $x_S \in Zq$ and computes the group public key, $y_S$, as, $y_S = g^{x_S} \mod p$.

(b). **SDC** assigns a public value $u_{S_i} \in Zq$ and secret value $k_i \in Zq, \forall\ i \in G_S$

(c). **SDC** constructs a $t^{th}$ degree polynomial $f_S(x)$ for each specified subset $H_S \in P(G_S)$, $|H_S| = t$, as,

$$f_S(x) = x_S \cdot \prod_{j \in H_S} \frac{(x - u_{S_j})}{(0 - u_{S_j})} + \sum_{i \in H_S} k_i \frac{x}{u_{S_i}} \cdot \left( \prod_{j \in H_S, J \neq i} \frac{(x - u_{S_j})}{(x - u_{S_j})} \right) \mod q,$$

with $x_S = f_S(0)$ and $f_S(u_{S_i}) = k_i, \forall\ i \in H_S$.

(d). **SDC** chooses a public value $u_{H_S} \in Zq$ for each specified subset $H_S \in P(G_S)$ such that $u_{H_S} \neq u_{S_i}, \forall\ i \in G_S$ and computes the specific verification key, $V_K$, as,

$$V_K = g^{f_a(u_{H_S}) \cdot \prod_{j \in H_S} \frac{(0 - u_{S_j})}{(u_{H_S} - u_{S_j})}} \mod p.$$

(e). **SDC** randomly selects $K \in Zq$ and publishes a value $W = g^{-K} \mod p$.

(f). **SDC** computes the secret shares $l_i, \forall\ i \in G_S$, as, $l_i = k_i \cdot \dfrac{-u_{H_S}}{(u_{S_i} - u_{H_S})} \mod q$.

(g). **SDC** publishes the public values $m_i$ and $v_{S_i}, \forall\ i \in G_S$, as,

$$m_i = g^{l_i} \mod p, \quad v_{S_i} = l_i \cdot y_{S_i}^{K} \mod p.$$

Here $y_{S_i}$ is the public key associated with each user $i$ in the group $G_S$.

## 7.2.2. Partial Signature generation by any specified subset $H_S$ of $t$ users and verification

The coordinator asks each members of the specified subgroup to sign the message $m$. If the members of a specified subset $H_S$ agree to sign, a message $m$ for a person $R$. $R$ possesses a pair $(x_R, y_R)$. Then the signature generation has the following steps.

(c) Each user $i \in H_S$ randomly selects $K_{i_1}$ and $K_{i_2} \in Z_q$ and computes

$$u_i = g^{-K_{i_2}} \mod p, \quad v_i = g^{K_{i_1}} \mod p \text{ and } w_i = g^{K_{i_1}} y_R^{K_{i_2}} \mod p.$$

(d) Each user makes $u_i$, $w_i$ publicly and $v_i$ secretly available to each member $i \in H_S$. Once all $u_i$, $v_i$ and $w_i$ are available, each member $i \in H_S$ computes the product $U_S$, $V_S$, $W_S$ and a hash value $R_S$, as,

$$U_S = \prod_{i \in H_S} u_i \mod q, \quad V_S = \prod_{i \in H_S} v_i \mod q,$$

$$W_S = \prod_{i \in H_S} w_i \mod q \text{ and } R_S = h(V_S, m) \mod q.$$

(c). Each user $i \in H_S$ recovers his/her secret share $l_i$, as, $l_i = v_{S_i} W^{x_{S_i}} \mod p$.

(d). Each user $i \in H_S$ modifies his/her shadow, as, $MS_{S_i} = l_i \cdot \prod_{j=1, j \neq i}^{t} \frac{-u_{S_j}}{u_{S_i} - u_{S_j}} \mod q$.

(e). Each user $i \in H_S$ uses his/her modified shadow and computes his partial signature $s_i$, as, $s_i = K_{i_1} + MS_{S_i} \cdot R_S \mod q$.

(f). Each member $H_S$ sends his partial signature to the **designated combiner** *DC*.

(g). The *DC* verifies the partial signature ($s_i$, $v_i$, $R_S$) by the congruence

$$g^{s_i} \stackrel{?}{\equiv} v_i \cdot m_i \left( \prod_{j \in H_S, j \neq i} \frac{-u_{S_j}}{u_{S_i} - u_{S_j}} \mod q \right)^{R_S} \mod p.$$

If this congruence holds, then the partial signature $(s_i, v_i, R_S)$ for shareholder $i$ is valid.

### 7.2.3. Group Signature Generation

(a). **DC** can computes the group signature $S_S = \sum_{i \in H_S} s_i \mod q$ by combining all the partial signature.

**Here we should note that there is no secret information associated with the DC.**

(b). **DC** sends $\{S_S, U_S, W_S, m\}$ to $R$ as signature of the **group $S$ on the message $m$.**

### 7.2.4. Signature verification by R

To verify the validity of the group signature $\{S_S, U_S, W_S, m\}$ **the verifier $R$ needs his/her secret key** $x_R$. This sub-section consists of the following steps.

(a). The verifier $R$ can recovers the values $R_R$ and $R_S$, as

$$R_R = W_S \cdot U_S^{x_R} \mod p \text{ and } R_S = h(R_R, m).$$

(b). The verifier $R$ uses the group public key $y_S$ and the corresponding specific verification key, $V_K$ and check the congruence $V_K^{R_S} g^{S_S} \stackrel{?}{\equiv} R_R \cdot y_S^{R_S} \mod p$. for the validity of the signature

If this congruence holds, then the group signature $\{S_S, U_S, W_S, m\}$ is valid signature of the organization $S$ on the message $m$.

### 7.2.5. Proof of validity by R to any third party C

This part of the protocol is same as in section 5.2.5

## 7.3. Security Discussions

In this sub-section, we discuss the security aspects of proposed **Generalized Directed - threshold– multisignature scheme**. Here we discuss several possible attacks. But show that, none of these can successfully break our system.

(a). **Can any one retrieve the secret keys $c = f_S(0)$ from the group public key $y_S$ ?**

This is as difficult as solving discrete logarithm problem. No one can get the secret key $c$, since $f_S$ is generated by the randomly and secretly selected parameters. On the other hand, by using the public keys $y_S$ no one can get the secret key $c$ because this is as difficult as solving discrete logarithm problem.

**(b). Can one retrieve the secret shares $l_i, \forall\ i \in G_S$, from the equation**

$$v_{S_i} = l_i \cdot y_{S_i}^{K} \bmod p\ ?$$

No because $K$ is a randomly and secretly selected integer by the SDC.

**(c). Can one retrieve the secret shares $l_i, \forall\ i \in G_S$, from the equation**

$$l_i = v_{S_i}\ W^{x_{S_i}} \bmod p\ ?$$

Only the user $i$ can recovers his secret shares $l_i$ because $x_{S_i}$ is secret key of the user $i \in G_S$.

**(d). Can one retrieve the modified shadow $MS_{S_i}$, integer $K_{i_1}$, $R_S$ and partial signature $s_i, i \in G_S$ from the equation $s_i = K_{i_1} + MS_{S_i} \cdot R_S \bmod q\ ?$**

It is computationally infeasible for a forger to collect the $MS_{S_i}$, integer $K_{i_1}$, $R_S$ and partial signature $s_i, i \in G_S$.

**(e). Can the designated combiner $DC$ retrieve the any partial information from the equation, $S_S = \sum_{i \in H_S} s_i \bmod q\ ?$**

Obviously, this is computationally infeasible for $DC$.

**(f). Can one impersonate a user $i \in H$ ?**

A forger may try to impersonate a user $i \in H_S$, by randomly selecting two integers $K_{i_1}$ and $K_{i_2} \in Z_q$ and broadcasting $u_i$, $v_i$ and $w_i$. But without knowing the secret shares, $l_i$ and $R_S$, it is difficult to generate a valid partial signature $s_i$ to satisfy the verification equations,

$$g^{S_i} \stackrel{?}{\equiv} v_i \cdot m_i \left( \prod_{j \in H_S, j \neq i} \frac{-u_{S_j}}{u_{S_i} - u_{S_j}} \bmod q \right)^{R_S} \bmod p.$$

(g). **Can one forge a signature $\{S_S, U_S, W_S, m\}$ by the following equation,**

$$V_K^{R_S} g^{S_S} \stackrel{?}{\equiv} R_R \cdot y_S^{R_S} \bmod p.$$

A forger may randomly selects an integer $R_R$ and then computes the hash value $R_S$ such that $R_S = h(R_R, m) \bmod q$. Obviously, to compute the integer $S_S$ is equivalent to solving the discrete logarithm problem. On the other hand, the forger can randomly select $R_S$ and $S_S$ first, then try to determine a value $R_R$ that satisfy the equation $V_K^{R_S} g^{S_S} \stackrel{?}{\equiv} R_R \cdot y_S^{R_S} \bmod p.$

However, according to the property of the hash function $h$, it is quite impossible. Thus, this attack will not be successful.

(h). **Can the members of a specified subset $H_S \in P(G_S)$, conspire to recover the polynomial $f_S(x)$?**

The secret polynomial $f_S$ can be reconstructed with the knowledge of any $(t + 1)$ pairs of $(u_{S_i}, f_S(u_{S_i}))$. When the members of any specified subset conspire to reconstructed the polynomial $f_S$, there is a need of one more pair, because $|H_S| = t$. However, to recover the extra secret share $f_S(u_{H_S})$ from $V_K$ is as difficult as solving the discrete logarithm problem. Thus, any $t$ or more shareholders cannot conspire to reconstruct the polynomial $f_S(x)$ by providing their own secret shares.

## Illustration

The following illustration is supporting our scheme for practical implementation. Suppose $|G_S| = (S_1, S_2, S_3, S_4, S_5)$ and $|H_S| = (S_1, S_3, S_5)$, $p = 47$, $q = 23$, $g = 6$.

**Group Secret Key and Secret Shares Generation for the organization $S$**

(a). **SDC** selects a group secret key $x_S = 18$ and computes $y_S = 9$.

(b). **SDC** randomly selects $K = 22$ and publishes a value $W = 6$.

(c). **SDC** assigns public values $u_{S_i}$ and secret values $k_i$ by the following table.

| Value \ Users | $S_1$ | $S_2$ | $S_3$ | $S_4$ | $S_5$ |
|---|---|---|---|---|---|
| $u_{S_i}$ | 11 | 9 | 14 | 8 | 17 |
| $k_i$ | 8 | 14 | 10 | 5 | 13 |

(d). **SDC** constructs a polynomial $f_S(x) = 22x^3 + 5x^2 + 17x + 18$ for specified subset $H_S$.

(e). **SDC** chooses $u_{H_S} = 10$ for specified subset $H_S$ and computes key $V_K = 2$.

(f). **SDC** computes the secret key, public key, secret share and public value for each member of the group $G_S$, as shown in the following table.

| USER \ VALUE | Secret $x_i$ | Public $y_i$ | Secret $k_i$ | Public $u_{S_i}$ | Secret $l_i$ | Public $m_i$ | Public $v_{S_i}$ |
|---|---|---|---|---|---|---|---|
| User–$S_1$ | 5 | 21 | 8 | 11 | 12 | 37 | 14 |
| User–$S_2$ | 15 | 2 | 14 | 9 | 2 | 36 | 1 |
| User–$S_3$ | 20 | 42 | 10 | 14 | 21 | 17 | 24 |
| User–$S_4$ | 4 | 27 | 5 | 8 | 2 | 36 | 14 |
| User–$S_5$ | 12 | 37 | 13 | 17 | 11 | 14 | 13 |

**Partial Signature generation by any specified subset $H_S$ of $t$ users**

If the members of a specified subset $|H_S| = (S_1, S_3, S_5)$ agree to sign, a message $m$ for a person $R$. $R$ possessing $x_R = 10$, $y_R = 18$, then the signature generation has the following steps.

(a). $S_1$ randomly selects $K_{1_1} = 7$, $K_{1_2} = 9$ and computes $u_1 = 16$, $v_1 = 4$ and $w_1 = 21$.

(b). $S_3$ randomly selects $K_{3_1} = 5, K_{3_2} = 11$ and computes $u_3 = 37, v_3 = 21$ and $w_3 = 1$.

(c). $S_5$ randomly selects $K_{5_1} = 9, K_{5_2} = 7$ and computes $u_5 = 12, v_5 = 3$ and $w_5 = 18$.

(d). Each user computes the product $U_S = 7, V_S = 17, W_S = 2$ and $R_S = 8$ (let).

(e). $S_1$ recovers his/her secret share $l_1 = 12$ and computes $MS_{S_1} = 13, s_1 = 19$.

(f). $S_3$ recovers his/her secret share $l_3 = 21$ and computes $MS_{S_3} = 16, s_3 = 18$.

(g). $S_5$ recovers his/her secret share $l_5 = 11$ and computes $MS_{S_5} = 20, s_5 = 8$.

### Partial signature verification and Signature generation by DC

(b) $DC$ verifies each partial signature. For example for user $S_2$,

$$s_3 = 18, v_3 = 21, m_3 = 17, R_S = 8, \prod_{j=1,5, i=2} \frac{-u_{S_j}}{u_{S_i} - u_{S_j}} \mod 23 = 15.$$

He checks $6^{18} \stackrel{?}{\equiv} 21 \cdot (17^{15})^8 \mod 47$. This holds. Similarly, he checks other partial signatures.

(b). $DC$ computes a group value $S_S = 22$ and sends $\{22, 7, 2, m\}$ as signature of the group $S$ for the message $m$.

### Signature verification by the person R

The verifier $R$ needs his/her secret key $x_R$ to verify the validity of the group signature $\{22, 7, 2, m\}$. This sub-section consists of the following steps: -

(a). The verifier $R$ can recovers the values $R_R = 17$ and $R_S = 8$.

(b). The verifier $R$ checks the congruence $2^8 \cdot 6^{22} \stackrel{?}{\equiv} 17 \cdot 9^8 \mod 47$ for the validity of the signature.

This congruence holds, so the group signature $\{22, 7, 2, m\}$ is valid signature of the organization $S$ on the message $m$.

### Proof of validity by R to any third party C

(a). $R$ computes $\mu = 32$ and $R_R = 17$ and sends $\{17, 22, 7, m, 32\}$ to C.

(b). C recovers $R_S = 7$ and check the concurrence $2^8 .6^{22} \overset{?}{\equiv} 17 . 9^8 \mod 47$ for the validity of the signature. This holds; so, C goes to the next steps.

(c). $R$ in a zero knowledge fashion proves to C that $\log_7 32 = \log_6 18$ as follows:-

- C chooses random $u = 22$, $v = 25$ and computes $w = 32$ and sends $w$ to $R$.
- $R$ chooses random $α = 27$ computes $β = 18$, $γ = 24$ and sends $β$, $γ$ to C.
- C sends $u, v$ to $R$, by which $R$ can verify that $w = 32$.
- $R$ sends $α$ to C, by which she can verify that $β = 18$ and $γ = 24$.

## 7.4. Remarks

In this chapter, we have proposed a **Generalized Directed –Threshold Multi - signature Scheme without *SDC*.** In this scheme,

- There is a *SDC* to generate group secret key and corresponding secret shares of shareholders.
- There is a **designated combiner *DC*** who takes the responsibility to collect and verify each partial signature and then produce a group signature, **but no secret information is associated with the *DC*.**
- Any malicious set of signers cannot impersonate any other set of signers to forge the signatures. In case of forgery, it is possible to trace the signing set.
- Any $t$ or more shareholders acting in collusion cannot conspire to reconstruct the group secret key by providing their secret shares.

# Appendix

## Summary of the papers published/under publication:

**No of papers published: 02.**

**No of papers under publication: 05.**

## Paper published:

1. ***Applications of Directed Signature Scheme,*** In South East Asian Journal of Mathematics and Mathematical Science-2 (1), p.p. 13 — 26.

2. ***A Directed signature scheme and its applications,*** In the proceeding of National conference on Information Security, **Sponsored by DRDO**, New Delhi -2003, p.p. 124 – 132.

## Paper under publication:

1. ***Threshold- Directed signature scheme.*** Communicated to Aligarh Math Bulletin.

2. ***Threshold Directed – Signature Scheme with Threshold Verification,*** Communicated to J. of Applied and Pure Mathematics, New Delhi.

3. ***Directed Threshold Multi – Signature Scheme,*** Communicated to J. of Natural Science Grukhul Khagri University Haridwar.

4. **The Generalized Directed Threshold Multi- Signature Scheme,** Communicated to GANIT SANDESH, J. of Rajasthan Ganit Parishad.

5. ***Directed Threshold Multi- Signature Scheme *without SDC*,*** Manuscript.

# References

# Papers

# Books

# Websites

| | | | |
|---|---|---|---|
| 1. | http://www.ams.org | 26. | http://www.freevbcode.com |
| 2. | http://www.austinlinks.com | 27. | http://www.ifca.com |
| 3. | http://www.bitpipe.com | 28. | http://www.inf.ethz.ch |
| 4. | http://www.bokler.com | 29. | http://www.iacr.org |
| 5. | http://www.bluesoft.com | 30. | http://www.ietf.org |
| 6. | http://www.cryptography.com | 31. | http://www.Krypt.cs.unisb.de |
| 7. | http://www.counterpane.com | 32. | http://www.jjtc.com |
| 8. | http://www.cryptography.org | 33. | http://www.maths.mq.edu.au |
| 9. | http://www.crypto.stanford.edu | 34. | http://www.mailru.com |
| 10. | http://www.cs.fsu.edu | 35. | http://www.missturtle.com |
| 11. | http://www.csa.com | 36. | http://www.newton.uor.edu |
| 12. | http://www.cs.umd.edu | 37. | http://www.pschit.monash.edu. |
| 13. | http://www.cesar.orul.gov | 38. | http://www.research.yale.com |
| 14. | http://www.caislab.icu.ac.kr | 39. | http://www.research.att.com |
| 15. | http://www.cryptoheaven.com | 40. | http://www.rsasecurity.com |
| 16. | http://www.certicom.com | 41. | http://www.rsa.org.uk |
| 17. | http://www.citeseer.nj.nec.com | 42. | http://www.r-s-a.org |
| 18. | http://www.codebreaker.dids.com | 43. | http://www.rsa.com |
| 19. | http://www.cryptix.org | 44. | http://www.std.com |
| 20. | http://www.domainmart.com | 45. | http://www.searchson.com |
| 21. | http://www.epic.com | 46. | http://www.simonsingh.com |
| 22. | http://www.efgh.com | 47. | http://www.siam.org |
| 23. | http://www.exp-math.uni-essen.de | 48. | http://www.springer.com |
| 24. | http://www.fraw.org.uk | 49. | http://www.xatrix.org |
| 25. | http://www.fitry.com | 50. | http://www.williamstallings.com |